\PassOptionsToPackage{
    protrusion=true,
    activate={true,nocompatibility},
    final,
    tracking=true,
    kerning=true,
    spacing=true,
    factor=1100}{microtype}
\documentclass[acmsmall,nonacm]{acmart}
\SetTracking{encoding={*}, shape=sc}{40}

\definecolor{lightblue}{rgb}{0.63, 0.74, 0.78}
\definecolor{seagreen}{rgb}{0.18, 0.42, 0.41}
\definecolor{orange}{rgb}{0.85, 0.55, 0.13}
\definecolor{silver}{rgb}{0.69, 0.67, 0.66}
\definecolor{rust}{rgb}{0.72, 0.26, 0.06}
\definecolor{purp}{RGB}{68, 14, 156}
\colorlet{darkrust}{rust!85!black}
\colorlet{darkseagreen}{seagreen!85!black}
\colorlet{darksilver}{silver!65!black}
\colorlet{darklightblue}{lightblue!65!black}
\usepackage{siunitx}
\DeclareSIUnit{\flop}{FLOP}    
\DeclareSIUnit{\TOPS}{TOPS}
\usepackage[nameinlink]{cleveref}  
\usepackage{etoolbox}
\usepackage{adjustbox}
\AtBeginEnvironment{tabular}{\footnotesize}   
\AtBeginEnvironment{tabular*}{\footnotesize}

\AtEndPreamble{\hypersetup{colorlinks=true}}

\begin{document}
\hypersetup{
  linkcolor=darkrust,
  citecolor=seagreen,
  urlcolor=darkrust,
}

\title{A predictive, per-engine energy roofline of the Apple Neural Engine}

\author{Spencer H. Bryngelson}
\orcid{0000-0003-1750-7265}
\affiliation{%
\department{Walker Department of Mechanical Engineering}
\department{Oden Institute for Computational Engineering and Sciences}
\institution{The University of Texas at Austin} \city{Austin} \state{Texas} \country{USA}}
\email{shb@utexas.edu}

\begin{abstract}
The Apple Neural Engine (ANE) is in over a billion devices, but is characterized indirectly via CoreML.
Using a CoreML-free fused-graph dispatch path, we measure the ANE, along with the Apple GPU and CPU on the same device, with per-engine energy accounting.
The consumer-available roofline does not predict the ANE in practice.
A critical correction is that the ANE sees two effective memory bandwidths: a slow activation path and a fast weight-stream path.
Two roofline ridges follow, and the weight-stream ridge is below that of the GPU.
The ANE single-operand compute peak decreases instead of plateauing.
Its weight stream is byte-bound in fp16 but limited by the dequantizer's element rate in int8.
As a result, int8 compression saves energy at unchanged throughput and an int4 lookup-table stream raises throughput and efficiency at a cost of 2.55 points of ImageNet top-1.
The corrected roofline identifies the lower-energy ANE paths and bounds it from below.
A constructed device map reproduces across four unique ANE generations.
Results show that operators should be placed by arithmetic intensity and fused for optimal performance.
\end{abstract}


\maketitle
\fancypagestyle{firstpagestyle}{%
  \fancyhf{}%
  \renewcommand{\headrulewidth}{0pt}%
  \renewcommand{\footrulewidth}{0pt}%
  \fancyfoot[C]{\footnotesize Accepted to \emph{ACM Transactions on Architecture and Code Optimization} (TACO), 2026.}}
\thispagestyle{firstpagestyle}

\section{Introduction}
\label{sec:intro}

The Apple Neural Engine (ANE) is among the most widely deployed machine-learning accelerators, present in every recent iPhone, iPad, and Apple-silicon Mac.
Despite this ubiquity, the engine's performance and energy profile has had little open measurement.
The publicly quoted figure is a peak 38 TOPS (tera-operations per second) for the M4 Neural Engine.
Community guidance, discussed in a community knowledge base~\cite{hollemans_neuralengine} and implicit in Apple's transformers-on-the-engine material~\cite{apple2022anetransformers}, states that the engine is worthwhile mainly for inference on large transformer models.
Both the principal figure and the large-model guidance come from CoreML, which is the only sanctioned interface to the ANE.
However, these figures instrument the hardware indirectly.
CoreML decides, by an opaque placement policy, which operators run on the ANE versus the GPU or CPU, and it exposes neither a fair per-operator dispatch nor energy accounting~\cite{ane_guide}.
The community's working assumptions about the ANE thus reflect a scheduler's defaults and optimization schemes, not the hardware envelope.

This work measures which workloads favor the ANE over the Apple GPU and CPU for the same device on speed and watts.
No CoreML-free method exists to dispatch arbitrary operators with controlled batch, dtype, and warm state.
We provide one with \texttt{aneforge}~\cite{aneforge_paper}, a fused-graph compiler and runtime that lowers a user's operator graph to a single fused ANE program and submits it through Apple's \texttt{e5rt} execution runtime.
Every timed operator runs on the ANE with controlled batch, data type, and warm state.
This gives a fair per-operator dispatch and a controlled device comparison.
The access mechanism is documented elsewhere~\cite{aneforge_paper,ane_guide}, though the instrument is described in full in the online supplement so the measurements can be independently judged.
This includes two checks that every benchmarked operator runs on the ANE and not a fallback device.
Against the ANE we run MLX, Apple's Metal-based array framework (fp16), for the GPU and numpy/Accelerate (fp32, via Apple's AMX matrix coprocessor) for the CPU, on an Apple M5 Pro (ANE architecture h17s; a 48~GB MacBook Pro, 18-core CPU and 20-core GPU).
The setup repeats the three-way roofline comparison, neural processing unit (NPU, or ANE here) versus GPU versus CPU, that was established for a datacenter TPU~\cite{jouppi2017tpu}, for an on-device NPU with energy measured throughout.
The fp32 Accelerate/AMX CPU path is a contextual reference, not a like-for-like fp16 comparand.
The main comparison is therefore the ANE against its companion GPU.

Prior characterizations reach the engine through CoreML or target other accelerators: smartphone-AI benchmarking~\cite{ignatov2019aibenchmark,reddi2020mlperf}, the per-engine Edge TPU evaluation \cite{seshadri2022edgetpu}, edge and mobile NPU energy studies~\cite{edgeai2024roofline,xu2025npullm,hybe2025,tummalapalli2026edge,laskaridis2024melting}, and a small body of direct-access ANE work \cite{kumaresan2026orion,singh2026anem4,yoon_ane_asahi}.
These works do not isolate the ANE in a per-engine energy roofline across operating regimes (see \Cref{sec:related} for further discussion).

Here, we present a corrected roofline.
The textbook single-roofline model mispredicts the ANE.
The primary correction is that the ANE has two effective bandwidths: a slow activation path and a fast weight-stream path, so two ridges, not one.
The weight-stream ridge that governs all compute-bound work lies below the GPU's, so an operator becomes compute-bound on the ANE at a lower arithmetic intensity than on the GPU.
Its weight stream is byte-bound in fp16 but limited by the dequantizer's element rate in int8, so what weight compression buys depends on the encoding and on the part.
Its single-fused-operand compute roof for general matrix--matrix multiplication (GEMM) falls instead of plateauing, so fusion, not a larger operand, keeps work on the fast tier.
We combine these three corrections into a per-engine energy roofline.
From a workload's arithmetic intensity it assigns each operator to the lower-energy engine and bounds its energy per operation, validated on held-out archetypes and a decode loop on the measured host (\Cref{sec:predictor}).
The regime-dependent device map that prior NPU-versus-GPU work reports follows from this two-ridge structure.
We test whether that map is a property of one part or of the engine family, and separate what reproduces across generations from what is specific to the measured host (\Cref{sec:three-gen}).

The measurement method ties idle-subtracted total-package energy to the roofline per engine (\Cref{sec:methodology}).
From it follow the three corrections (\Cref{sec:weight-stream,sec:compute-roof,sec:decode}), a closed-form energy predictor and placement rule fit from two per-device constants (\Cref{sec:predictor}), and the device map's reproduction across four ANE generations (\Cref{sec:device-map,sec:three-gen}).

\section{Methodology}
\label{sec:methodology}

\paragraph{Measurement tool and dtype.} The ANE is driven by \texttt{aneforge}'s fused-graph dispatch (fp16): each operator graph is lowered to one fused MIL program and submitted directly to the engine, with no CoreML model and therefore no CoreML placement policy.
Every timed operator is verified to run on the ANE rail.
The per-rail power attribution below confirms this, and a route-and-capability gate test compiles both lowerings of every selectable operator on-device and validates each against a reference.
The full instrument description is in the supplemental material.
The GPU is driven by MLX~v0.31.2 (Metal, fp16; v0.32.2 for the compiled MiniLM and ViT-B/16 baselines).
The real-model rows also use torch-MPS (torch's Metal Performance Shaders backend) in fp16, with a single-session fp32 energy column.
The CPU is driven by numpy/Accelerate (fp32, AMX), with torch as the reference where it is the canonical implementation.
The GPU and ANE therefore run the same fp16 math; the CPU runs fp32 because Accelerate/AMX has no fast fp16 GEMM (it upcasts), which doubles its bytes-per-element and halves its arithmetic intensity, carried through the roofline arithmetic wherever it matters.

\paragraph{Speed.} The measured latency is the minimum end-to-end latency over repetitions after warmup, a dedicated probe separate from the sustained power window; at the smallest shapes it measures each device's dispatch floor.
Each repetition forces device completion (\texttt{mx.eval} for MLX, the compiled-net call for the ANE, \texttt{torch.mps.synchronize} for MPS) so that no work is left in flight.
All measurements are at warm caches and steady state, with matched batch sizes.
Throughput is GFLOP/s where a closed-form FLOP count exists, otherwise elements/s.
The latency includes each device's host and dispatch overhead; at sub-millisecond shapes the overhead is part of the cost a deployment pays and is not subtracted out.

\paragraph{Power.} Power is read from \texttt{powermetrics} per ANE, GPU, or CPU rail.
The state of programmatic Apple-silicon power measurement (IOReport versus \texttt{powermetrics}, and the absence of a validated estimator) is the subject of \cite{mlenergy_macprofiling}; prior Apple-silicon efficiency studies have read \texttt{powermetrics} per-rail across the M1 to M4 generations \cite{apple_power}.
We adopt total-package active power herein.

Following \cite{fanariotis2025npumcu}, a no-workload baseline is sampled once per run and subtracted (package \qty{1196}{\milli\watt} in the device-map run).
We take \texttt{powermetrics}' own \texttt{Combined Power (CPU + GPU + ANE)} line, not a hand-summed rail total, because an ANE workload still draws from host CPU power on dispatch (as does an MLX workload).

\paragraph{The measurement boundary.} By \emph{per-engine energy accounting} we mean that every operating point reports total-package power, latency, and accuracy for the engine that ran it.
The claim is one of attribution, not of an exhaustive power domain.
The domain is the CPU, GPU, and ANE rails as \texttt{powermetrics} reports them.
It excludes the DRAM rail, quantified next, and rests on a modeled SoC estimator, not a wall-socket meter.
For each measurement a per-rail breakdown is reported.
Median loaded package power with the coefficient of variation (CV\%) over per-sample readings are included, with CV above \qty{35}{\percent} flagged low-confidence measurement.
Each loop drives the workload for the entire powermetrics window and samples at \qty{500}{\milli\second} to integrate the ANE's sub-millisecond duty cycle; a naive fixed-time stop integrating at \qty{100}{\milli\second} produces a \qtyrange{60}{140}{\percent} CV quantity, whereas this strategy gives \qtyrange{1}{25}{\percent} CV (30 of 31 measurements below \qty{17}{\percent}).

\paragraph{Cross-validation against an independent estimator.} The total-package active-power metric is cross-validated against Apple's IOReport energy-model counters at five operating points spanning the attribution axis (see supplement, Table~S1), where the estimators agree to a median of \qtyrange{97}{99}{\percent}.
Both remain modeled SoC estimators, not wall-socket meters; the relative device comparisons are unaffected by a shared offset that cancels via ratios.

\paragraph{The DRAM rail.} The combined power excludes the DRAM rail, which the IOReport DRAM channel puts at \qtyrange{2.4}{12.9}{\percent} of package across the loaded operating points.
That share is device-dependent, not uniform: \qtyrange{4.1}{5.3}{\percent} on an ANE convolution against \qtyrange{12.5}{12.9}{\percent} on a saturating GPU GEMM.
Including DRAM takes the GPU-over-ANE package-power ratio on that pair from $8.4$ to $9.0$.
Since the GPU's DRAM share is the largest measured, excluding the rail is the conservative choice for the ANE-versus-GPU comparisons this paper draws (online supplement, Table~S28).

\paragraph{Derived metrics: single-window.} Perf/watt and mJ/inference (active W $\times$ sustained ms/iter) are computed within a single sustained measurement, never spliced across executions.
The sustained per-iteration time is the wall time of the measurement loop divided by the inferences it completed inside the same window over which power is summarized.
Each \qty{6}{\second} window holds 354 to \num{61446} inferences, so the reported energy integrates completed work within one execution, and does not combine statistics from separate runs.
At sub-millisecond timings that time exceeds the separate minimum-latency probe on every ANE path, and perf/watt therefore cannot be reconstructed from the latency column, by design.
Across \Cref{tab:device-map} the two statistics differ by a median factor of $1.05$, and by more than \qty{25}{\percent} only at the smallest shapes, where the gap measures per-call dispatch overhead.
The online supplement (Table~S21) reports both per class, with the sensitivity of each energy figure to the choice of power and latency statistic; both choices are conservative for the ANE.

\paragraph{Repeated measurements.} Some results are replicated over independent cold starts, each a fresh process with its own idle baseline and randomized device order, and carry a \qty{95}{\percent} confidence interval.
These are the ANE and fp16 torch-MPS energies and ratios in \Cref{tab:real-models}, the six device-map classes marked in \Cref{tab:device-map}, and the ceiling and serving comparisons and crossovers of Table~S17 in the online supplement.
The compiled-MLX columns of \Cref{tab:real-models} are repetitions within one session, also with \qty{95}{\percent} intervals.
The other seven device-map classes, the fp32 GPU energy column, the serving sweeps apart from their crossovers, the decode sweeps, and the cross-generation maps are each a single sustained measurement with a within-window CV and no across-run interval.
Under repetition the DFT ratio moved from its single-session $0.82\times$ to $1.35\times$, with an interval spanning parity (\Cref{sec:device-map}).

\paragraph{Accuracy.} Every workload reports relative L2 error against an fp32/fp64 numpy reference, printed beside the speed and watt numbers so that a speed gain at worse accuracy is visible.
The GPU is consistently more accurate in fp16 than the ANE, most so on large GEMMs.
This error is computed on a synthetic input and is not a uniform proxy for real data: on their own domains ResNet-18 falls to \num{1.6e-3}, MiniLM holds at \num{5.7e-3}, and ViT-B/16 rises to \num{2.6e-2}.
Compiling the ViT encoder to each depth localizes that gap.
Error grows by \qtyrange{7}{9}{\percent} per block for both inputs, and their difference appears only at the classification head, where a peaked logit vector carries larger absolute rounding over a smaller norm.
Relative L2 also does not say whether a divergence changes an answer, so all three real models are scored on their task.
ResNet-18 reaches \qty{69.76}{\percent} top-1 on the ANE, the GPU, and the fp32 CPU reference alike over all \num{50000} ILSVRC2012 validation images.
ViT-B/16 reaches \qty{81.00}{\percent} against the reference's \qty{81.06}{\percent}.
Both reproduce the published torchvision accuracies.
Neither ANE path shifts accuracy: its label changes run in both directions (ResNet 14 against 14, ViT 33 against 21, McNemar $p=1.00$ and $0.13$), though on ViT it changes ten times as many individual labels as the GPU does.
MiniLM has no label set, so the decision measured there is which neighbor an embedding retrieves: the ANE returns the fp32 nearest neighbor for \qty{99.2}{\percent} of queries and reproduces the pairwise similarity ranking exactly to four decimals.
Weight compression produces the measurable task-level losses: per-channel int8 costs \qty{0.11}{} percentage points of ResNet-18 top-1 and two points of MiniLM retrieval agreement, and int4 costs \qty{2.55}{} points (online supplement, Tables~S25 and~S26).

\paragraph{Exclusions.} Sixteen subprocess-bridge operators (sort, cost-volume, farthest-point sampling, radius search, and others) run at \qty{25}{\milli\second} to \qty{2.5}{\second} because each call spawns a one-shot probe process and pays the full process-spawn, graph-author, and ANE-load cost every call, a dispatch artifact, not silicon speed.
Timing them against MLX would misrepresent the hardware, so they are excluded from the speed comparison and treated as a capability statement instead.
The sixteen are listed in the online supplement, and the three bridge operators that carry a persistent worker (\texttt{sdpa}, \texttt{argmax}, \texttt{topk}) are timed on silicon and remain in the comparison.

\section{Compute and bandwidth ceilings}
\label{sec:ceilings}

The corrected roofline uses measured ceilings for each device.
A representative shape need not saturate the device: a $512\times512$ GEMM reaches only about \qty{4}{\percent} of the GPU's fp16 peak and therefore understates its capability.

\subsection{Compute roof}
\label{sec:compute-roof}

The compute-saturation sweep scales two compute-bound primitives (square GEMM, $2N^3$ FLOPs, and a $3\times3$ same-pad convolution, the ANE's highest-throughput primitive) until each device plateaus, measuring throughput, idle-subtracted active package power, and fp16 relative error at every size.
\Cref{tab:compute-peaks} gives the peaks.

\begin{table}[htbp]
  \centering
  \small
  \caption{Compute and efficiency peaks (M5 Pro), GPU and ANE in fp16.}
  \label{tab:compute-peaks}
  \begin{tabular}{lllrlrr}
    \toprule
    Primitive & Device & Dtype & Peak GFLOP/s & @ size & relerr there & Peak GFLOP/s/W \\
    \midrule
    GEMM & \textbf{GPU} & fp16 & \textbf{30\,862} & $N=6144$ & 3.6e-4 & 603 ($N=8192$) \\
    GEMM & \textbf{ANE} & fp16 & 10\,191 & $N=2048$ & 1.2e-2 & \textbf{866} ($N=4096$) \\
    conv & \textbf{GPU} & fp16 & \textbf{31\,790} & C=512 & 7.4e-3 & 1\,392 (C=512) \\
    conv & \textbf{ANE} & fp16 & 18\,771 & C=512 & 5.5e-4 & 1\,372 (C=512) \\
    \bottomrule
  \end{tabular}
\end{table}

The GPU has the higher raw throughput on both primitives compared to the other devices; GEMM $3.0\times$ the ANE (30.9 versus \qty{10.2}{\tera\flop\per\second}) and conv $1.7\times$ the ANE (31.8 versus \qty{18.8}{\tera\flop\per\second}), confirming that the large square GEMM of large-language-model (LLM) inference is appropriate for the same SoC's GPU.
Perf/watt divides between the primitives.
At $N=4096$ the ANE is $3.6\times$ slower but draws \qty{9.5}{\watt} against the GPU's \qty{53.3}{\watt}, so its efficiency at that size is $1.55\times$ the GPU's (866 against 558 GFLOP/s/W).
Peak against peak, the ANE's efficiency is $1.4\times$ higher (866 at $N=4096$ against the GPU's 603 at $N=8192$).
Cold-start replication of the device map's GEMM-compute shape ($M=256$, $K=N=4096$) gives $1.45\times$ (811 against 558 GFLOP/s/W, online supplement, Table~S17); this is the most variable contested claim, but it remains Holm-significant.
For conv the single-sweep gap is small (ANE 1372 against GPU 1392 GFLOP/s/W) and replication resolves it to a \qty{5}{\percent} GPU edge.
This bounds the efficiency claim: it holds for GEMM at peak and reverses to a small GPU edge at conv saturation.
The utilization shows in the power curves: the GPU's rises from \qty{9}{\watt} at $N=512$ and plateaus at \qtyrange{49}{53}{\watt} for $N \geq 4096$, while the ANE's peaks at \qty{12.7}{\watt} at $N=2048$ and then falls with its throughput.

\begin{figure}[t]
  \centering
  \includegraphics{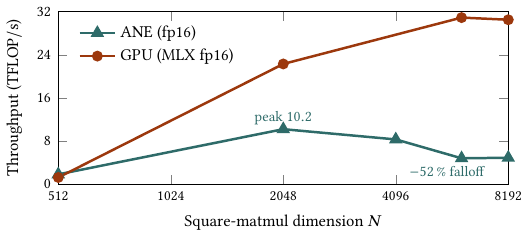}
  \Description{Square-matmul throughput versus dimension for the ANE and GPU; the ANE peaks near 10.2 TFLOP/s at $N=2048$ then falls off by 52 percent, while the GPU rises and plateaus above 30 TFLOP/s.}
  \caption{Square-matmul throughput versus dimension $N$ on the M5 Pro (ANE fp16, GPU MLX fp16).
The ANE roof peaks then falls; the GPU's rises and plateaus.}
  \label{fig:gemm-falloff}
\end{figure}

The second roofline correction is determined via the throughput curve: the ANE's compute roof falls instead of plateauing.
ANE GEMM peaks at $N=2048$ (\qty{10.2}{\tera\flop\per\second}), then degrades to \qty{8.3}{\tera\flop\per\second} at $N=4096$, 4.8 for $N=6144$, and 4.9 for $N=8192$, a \qty{52}{\percent} drop that reproduces across runs (\Cref{fig:gemm-falloff}), remaining at roughly half its peak for $N \geq 6144$.
The active power falls with the throughput (\qty{12.7}{\watt} to \qty{6}{\watt}), so the engine is stalled, not busy at a reduced rate.
Once a square operand exceeds the ANE's working-set window (about $N=2048$ to 4096), on-chip tiling and weight streaming no longer feed the multiply--accumulate (MAC) array fully.
The computation then serializes on off-chip streaming.
The same peak-then-decline shape appears on an M2 Pro (h14), peaking at \qty{7.5}{\tera\flop\per\second} at $N=2048$ and falling to \qty{2.2}{} at $N=8192$, so it is a property of the engine, not of one part.
Accuracy degrades over the same range: relative error grows from \num{2.6e-3} at $N=512$ to \num{6.2e-2} at $N=8192$.
Large-$N$ ANE GEMM therefore loses both throughput and accuracy.
That growth is consistent with per-product fp16 rounding over a long contraction, not a narrow accumulator: the MAC array carries a wide, fp32-class accumulator, and the error tracks $\sqrt{K}$ as independent rounding errors accumulate.

The falloff is specific to the large square single-matmul shape: ANE conv throughput rises monotonically to its \qty{18.8}{\tera\flop\per\second} peak with relative error constant at \num{5e-4}, because conv's natural tiling keeps the array fed.
This is why conv, not GEMM, is the ANE's highest-throughput primitive.
A check robust to the falloff isolates the raw ceiling: deepening a fused chain of small matmuls grows compute at fixed I/O, and its slope reaches \qty{19.6}{\tera\flop\per\second} against the \qty{10.2}{} single-matmul peak, so the gap is utilization, not a silicon limit.

\subsection{Bandwidth roof}
\label{sec:bandwidth-roof}

The compute-bound primitives (matmul, conv, and their variants) are the minority of \texttt{aneforge}'s 50 fused operators.
The rest (activations, elementwise binaries, reductions, softmax, the norm family, pooling, and data-movement ops) are memory-bandwidth-bound at size.
We scale each class from $256$K to $64$M elements and read its saturation plateau; the rise toward that plateau identifies the dispatch-bound small sizes.
\Cref{tab:bw-ceiling} shows the ceilings.

\begin{table}[htbp]
  \centering
  \small
  \caption{Bandwidth ceilings (M5 Pro), GPU and ANE in fp16.}
  \label{tab:bw-ceiling}
  \begin{tabular}{lrrrr}
    \toprule
    Archetype (bytes) & ANE GB/s & GPU GB/s & ANE GB/s/W & GPU GB/s/W \\
    \midrule
    \textbf{streaming} relu / x$\cdot$2 (2N) & 24.1 & \textbf{229.7} & 1.99 & \textbf{27.13} \\
    reduction sum (1N) & 22.4 & \textbf{193.0} & 2.24 & \textbf{27.28} \\
    softmax (2N) & 15.1 & \textbf{238.4} & 1.78 & \textbf{18.54} \\
    layer\_norm (2N) & 4.4 & \textbf{41.9} & 0.90 & \textbf{2.43} \\
    gelu light-compute (2N) & 23.2 & 48.6 & 2.03 & 3.70 \\
    \bottomrule
  \end{tabular}
\end{table}

The streaming row sets the measured GPU ceiling at \qty{230}{\giga\byte\per\second}, with softmax reaching \qty{238}{}.
This single-program rate is about \qty{75}{\percent} of the part's \qty{307}{\giga\byte\per\second} unified-memory specification and is the appropriate ceiling for a measured roofline.
The ANE's achieved bandwidth through \texttt{aneforge}'s single-fused-program path is flat at \qtyrange{22}{24}{\giga\byte\per\second} across streaming and reduction, reached already at the smallest sizes measured, about $10\times$ below the memory ceiling the GPU attains.
That flat-from-small signature is the finding: on pure elementwise and reduction graphs the ANE is bound by its program-streaming and dispatch path, not by the memory system, so the \qty{24}{\giga\byte\per\second} is a path ceiling and not a silicon-memory one.
The GPU therefore has both the higher raw bandwidth (230 against 24) and the higher bandwidth-per-watt on every archetype (27.1 against \qty{1.99}{\giga\byte\per\second\per\watt} on streaming).
The ANE's usual perf/watt advantage does not survive here, because it came from doing useful compute at low power and a pure memory stream is the one regime with no compute to do.
The consequence is that standalone large softmax, elementwise, and reduction ops belong on the GPU; they are cheap only when fused into a larger compiled program.

\subsection{Weight-stream bandwidth (GEMV-K)}
\label{sec:weight-stream}

\begin{figure}[t]
  \centering
  \includegraphics{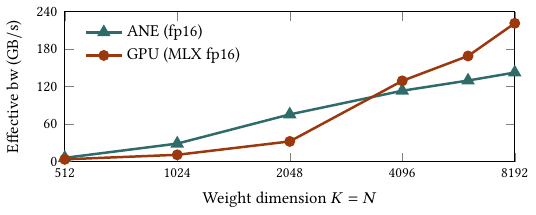}
  \Description{Effective weight-stream bandwidth versus weight dimension K for a GEMV on the ANE and GPU; the ANE is higher through $K=2048$ and the GPU is higher from $K=4096$.}
  \caption{Weight-stream effective bandwidth versus weight dimension $K=N$ for a decode GEMV ($M=1$) on the M5 Pro.
  The ANE's bandwidth exceeds the GPU's up to $K=2048$ (\qty{76}{\giga\byte\per\second} versus \qty{32}{\giga\byte\per\second}) as it amortizes its dispatch floor, then plateaus near \qtyrange{130}{143}{\giga\byte\per\second}; the GPU's unified memory is higher from $K=4096$ and reaches \qty{221}{\giga\byte\per\second}.}
  \label{fig:weight-stream}
\end{figure}

This measurement establishes the first roofline correction: the ANE has two effective bandwidths, not one.
The streaming roofline of \Cref{sec:bandwidth-roof} measures the standalone-activation path.
A decode GEMV (matrix--vector multiply, $M=1$) exercises a different path: it streams a $K \times N$ weight matrix once and does negligible compute (arithmetic intensity AI $\approx 1.0$), so its effective bandwidth is that of the path the ANE uses to stream weights, not of its standalone-activation path.
A single GEMV at $K=N=4096$ gives an effective \qty{112}{\giga\byte\per\second}, well above the \qty{24}{\giga\byte\per\second} activation ceiling of \Cref{sec:bandwidth-roof}.
To resolve whether this is anomalous or a second bandwidth tier, we sweep $K=N \in \{512,1024,2048,4096,6144,8192\}$ and read effective GB/s = (weight + I/O bytes, fp16) / minimum latency (the per-archetype byte models are in the online supplement, Table~S16).
\Cref{fig:weight-stream} reports the sweep (the per-$K$ values are tabulated in the online supplement, Table~S3); the same rise-to-plateau standard we applied to the \qty{24}{\giga\byte\per\second} figure applies here.

The ANE's bandwidth increases 6 $\to$ 29 $\to$ \qty{76}{\giga\byte\per\second} at small $K$, and so is dispatch-bound and the flat-from-small signature of \Cref{sec:bandwidth-roof} is inverted: here the per-program overhead is amortized as the weight matrix grows (\Cref{fig:weight-stream}).
The ANE reaches \qty{113.5}{\giga\byte\per\second} for $K=4096$, reproducing the \qty{112}{\giga\byte\per\second} figure to within \qty{2}{\percent}, then plateaus to \qtyrange{130}{143}{\giga\byte\per\second} at larger $K$.
The 112 figure is thus a conservative lower bound, and the weight-stream path is 5 to $6\times$ the \qty{24}{\giga\byte\per\second} standalone path.

The standalone-activation path plateaus at \qty{23}{\giga\byte\per\second} with a stable large-size ceiling (a plateau coefficient of variation of \qty{3.3}{\percent}) and does not rise with size, so it is a true ceiling and not an under-amortized operating point of the same DMA; the weight-stream tier is $5.4\times$ higher.
Fragmenting a fixed weight volume into many small programs also lowers the achieved weight-stream bandwidth, consistent with a separate weight-stream tier.
An M2 Pro (h14, a desktop Mac mini) shows the same flat low activation ceiling at \qty{14}{\giga\byte\per\second} and a weight-stream tier $3.8\times$ higher; an M4 (h16, a second Mac mini) shows the same flat ceiling near \qty{10}{\giga\byte\per\second} with a weight-stream path that runs $2.4$ to $6.6\times$ above it.

\paragraph{Scope of the two-tier measurement.} The two effective bandwidths are a property of the dispatch path as reached by \texttt{aneforge}, measured end to end from the host, and the measurement does not establish a dedicated weight DMA as the mechanism.
In these programs the weight matrix is a compile-time constant of the compiled graph and the activation is the only runtime input, so a warm repetition reissues the same program against the same constant.
The faster tier is also consistent with constant handling, on-chip tensor layout, compiler placement, or residency across repetitions.
An external measurement cannot separate these from a distinct transfer engine, so we report a dedicated weight path as a candidate mechanism, not an established one.
Every weight point exceeds the activation ceiling by at least $2.4\times$, so the two-tier separation holds across sizes; the online supplement notes one h16-only weight-path stride-aliasing dip at exact powers of two, which does not affect the separation.
The activation path's saturation at \qty{14}{\giga\byte\per\second} on a part with roughly \qty{200}{\giga\byte\per\second} of unified-memory bandwidth indicates the ceiling is a dispatch-path limit, not a DRAM limit.
The absolute ceilings track the part, not the generation: the two Pro parts (23 and \qty{14}{\giga\byte\per\second}) sit above the smaller base M4 (\qty{10}{\giga\byte\per\second}), even though the M4 is the newer silicon.

The ANE's single-GEMV bandwidth exceeds the GPU's through $K=2048$ (29 versus 11 at $K=1024$, 76 versus 32 at $K=2048$); from $K=4096$ the GPU's unified memory is higher, rising 129 $\to$ \qty{221}{\giga\byte\per\second} while the ANE saturates near 143.
(The CPU's fp32 figures are highest only at the smallest sizes, where the tiny weight fits in cache and the comparison is call-overhead-bound, not bandwidth-bound; it falls to \qtyrange{50}{60}{\giga\byte\per\second} once the matrix spills.)

The energy column shows the ANE streams weights at higher GB/s/W than the GPU at every $K$, $2.0\times$ at $K=4096$ (35.9 versus 18.0) and still $1.45\times$ at $K=8192$ (60.8 versus 41.9), past the bandwidth crossover.
Even where the GPU moves more bytes per second, the ANE moves them at lower energy on the exact path decode uses, which is the roofline-level reason the ANE's decode advantage is energy (\Cref{sec:decode}).

The weight-stream path also accepts compressed weights, decompressed during the DMA (the on-the-fly-decode pattern of EIE \cite{han2016eie}); \texttt{aneforge} emits int4 lookup-table palettization and unstructured sparse forms behind one opt-in flag.
On a single matmul this raises arithmetic intensity and speeds the stream ($1.65$ to $1.82\times$ on large layers, online supplement Table~S4), moving the int4 operating point to \qty{104}{\percent} of the matmul roof at $B=128$ (\Cref{tab:roofline-place}).
End to end, int8's benefit is footprint and energy, not latency: the int8 weight stream is bound by the dequantizer's element rate, not by bytes, the third roofline correction, developed in \Cref{sec:decode} (halving the weight bytes yields energy savings on a part at its element-rate ceiling, and throughput only on a bandwidth-starved one, though a lookup-table encoding raises the ceiling itself).

\section{Per-device roofline synthesis}
\label{sec:roofline}

The corrected roofline turns on one inequality: the ANE's weight-stream ridge (about \qty{71}{\flop\per\byte}) lies below the GPU's (134), so the path that carries every compute-bound ANE operator becomes compute-bound at a lower arithmetic intensity than the GPU does.

\begin{figure}[t]
  \centering
  \includegraphics{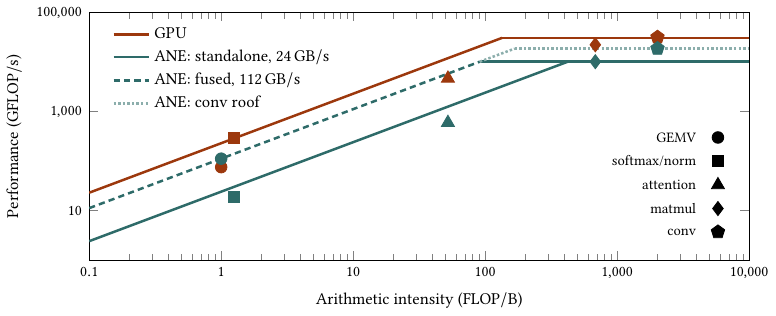}
  \Description{Per-device roofline plot of performance versus arithmetic intensity for the ANE and GPU, with five operator archetypes measured on both devices; the ANE shows two bandwidth slopes (24 and 112 GB/s) and the GPU is higher on every compute-bound archetype except decode GEMV.}
  \caption{Per-device roofline on the M5 Pro: five operator archetypes on the ANE (triangles) and the GPU (circles), placed by arithmetic intensity.
The ANE shows two bandwidth slopes, a \qty{24}{\giga\byte\per\second} standalone-activation path and a faster weight-stream path used by compiled programs; per-archetype, per-device placement is in \Cref{tab:roofline-ridge,tab:roofline-place}.}
  \label{fig:roofline}
\end{figure}

A roofline positions each operator by its arithmetic intensity (AI = FLOPs / bytes-moved) and predicts an achievable ceiling of \texttt{min(compute\_roof,}\ \texttt{bandwidth\_roof}\,$\times$\,\texttt{AI)}; the ridge point (\texttt{compute\_roof}\,/\,\texttt{bandwidth\_roof}) divides memory-bound (left) from compute-bound (right).
\Cref{tab:roofline-ridge} gives each device's roofs and ridges (recomputed from the two ceiling JSONs, not re-measured); \Cref{tab:roofline-place} then tabulates every measured archetype by its arithmetic intensity, reporting its achieved rate and the fraction of the applicable roof it reaches on each device.
Only the high-AI matmul and conv archetypes exceed any device's ridge (compute-bound); everything below falls on the bandwidth slope.

\begin{table}[htbp]
  \centering
  \small
  \caption{Per-device roof and ridge (recomputed from the saturation and bandwidth JSONs).
  The ANE has two bandwidths (\Cref{sec:bandwidth-roof,sec:weight-stream}) and therefore two ridges: a standalone-activation path near \qty{24}{\giga\byte\per\second} and a weight-stream path near \qty{143}{\giga\byte\per\second}.
}
  \label{tab:roofline-ridge}
  \begin{tabular}{llrrr}
    \toprule
    Device & Dtype & Compute roof (GFLOP/s) & Bandwidth roof (GB/s) & Ridge (FLOP/byte) \\
    \midrule
    GPU & fp16 & gemm \textbf{30862} / conv 31790 & 229.7 & \textbf{134.4} \\
    ANE (activation path) & fp16 & gemm \textbf{10191} / conv 18771 & 24.1 & \textbf{423.6} \\
    ANE (weight-stream path) & fp16 & gemm \textbf{10191} / conv 18771 & $\sim$143 & \textbf{$\sim$71} \\
    \bottomrule
  \end{tabular}
\end{table}

\begin{table}[htbp]
  \centering
  \small
  \caption{Operator archetypes placed on each device's roofline: arithmetic intensity (AI, the fp16 value), achieved performance, and the fraction of the applicable roof (\texttt{min(compute,}\ \texttt{bandwidth}\,$\times$\,\texttt{AI)}) reached, M5 Pro.
  Only the high-AI matmul and conv rows exceed any ridge (compute-bound); the rest fall on the bandwidth slope.
  $^\ddagger$The ANE GEMV is scored against the weight-path roof (\qty{78}{\percent}), not the \qty{24}{\giga\byte\per\second} activation roof (a spurious \qty{466}{\percent}), since a weight-streaming GEMV uses the weight path.}
  \label{tab:roofline-place}
  \begin{tabular}{lrrr}
    \toprule
    Operator archetype & AI & GPU GFLOP/s (\%roof) & ANE GFLOP/s (\%roof) \\
    \midrule
    matmul $N=512$ & 171 & 1235 (\qty{4}{\percent}) & 1819 (\qty{18}{\percent}) \\
    matmul $N=2048$ & 683 & 22345 (\qty{72}{\percent}) & 10191 (\qty{100}{\percent}) \\
    matmul $N=8192$ & 2731 & 30527 (\qty{99}{\percent}) & 4882 (\qty{48}{\percent}) \\
    conv ($C=64$) & 287 & 5343 (\qty{17}{\percent}) & 5885 (\qty{31}{\percent}) \\
    conv ($C=512$) & 2020 & 31790 (\qty{100}{\percent}) & 18771 (\qty{100}{\percent}) \\
    attention ($S=512$) & 52 & 4685 (\qty{39}{\percent}) & 598 (\qty{48}{\percent}) \\
    GEMV / decode & 1.0 & 75 (\qty{33}{\percent}) & 112 (\qty{78}{\percent})$^\ddagger$ \\
    softmax & 1.25 & 298 (\qty{104}{\percent}) & 19 (\qty{63}{\percent}) \\
    gelu & 2.5 & 122 (\qty{21}{\percent}) & 58 (\qty{96}{\percent}) \\
    layer\_norm & 2.5 & 105 (\qty{18}{\percent}) & 11 (\qty{18}{\percent}) \\
    relu / copy & 0.25 & 57 (\qty{100}{\percent}) & 6 (\qty{100}{\percent}) \\
    \midrule
    matmul int4 ($4096^2$, $B=1$) & 4 & 171 (\qty{19}{\percent}) & 165 (\qty{29}{\percent}) \\
    matmul int4 ($4096^2$, $B=128$) & 410 & 5218 (\qty{17}{\percent}) & 10579 (\qty{104}{\percent}) \\
    \bottomrule
  \end{tabular}
\end{table}

On its standalone-activation path the ridge is about \qty{424}{\flop\per\byte} ($3.2\times$ the GPU's 134), a consequence of that path's bandwidth being only \qty{24}{\giga\byte\per\second} against \qty{10}{\tera\flop\per\second} of compute.
This is why dispatching an elementwise, norm, or softmax op one at a time underutilizes the engine: such ops lie far to the left of 424 and never approach compute-bound.

Compute-bound work (GEMM, conv, GEMV, linear, anything that streams weights) instead uses the weight-stream path (\Cref{sec:weight-stream}), whose ridge is only about \qty{71}{\flop\per\byte}, below the GPU's 134.
On the path that carries the ANE's compute-bound work, an operator becomes compute-bound at a lower arithmetic intensity than on the GPU, not higher.
This marks where the ANE's energy advantage begins, not which device is faster: throughput follows the lower, falling compute roof (\Cref{sec:compute-roof}), so on large square GEMM the GPU is the faster device while the ANE still draws far less power, whereas on conv and the stencil the ANE is faster and lower-power.
A single ridge at 424 conflates the activation and weight paths and places the weight-streaming points above the roof.
With both ridges, the GEMV reaches \qty{78}{\percent} of its applicable weight-path roof instead of the \qty{466}{\percent} the activation ridge implies.
The CPU (fp32 via Accelerate/AMX) is a contextual reference whose full figures are in the online supplement (Table~S18); the like-for-like comparison is GPU versus ANE, which run identical fp16 math.

The fused 3-conv block at about \qty{110}{\percent} of the single-conv roof exceeds a standalone-dispatch ceiling by amortizing per-dispatch overhead and I/O (the fusion discussion below).
The GPU softmax at \qty{104}{\percent} is FLOP-convention sensitivity (the flat per-element transcendental allowance of the online supplement).

The ANE roofline carries two caveats.
It is not a single clean line: its compute roof differs by operator (conv about 18.8 versus square-matmul about \qty{10.2}{\tera\flop\per\second}, and matmul falls off past $N=2048$ to about \qty{4.9}{\tera\flop\per\second} at $N=8192$), so we draw both roofs.
The \qty{24}{\giga\byte\per\second} bandwidth and the ridge near \qty{424}{\flop\per\byte} are also a path operating point of \texttt{aneforge}'s single-program streaming, not a silicon DMA limit (\Cref{sec:bandwidth-roof}); the weight-stream path reaches 5 to $6\times$ higher at its plateau (\Cref{sec:weight-stream}).
On the figure we draw the weight-stream line at the conservative \qty{112}{\giga\byte\per\second} operating point corroborated at $K=4096$ (ridge about \qty{91}{\flop\per\byte}); the measured plateau reaches about \qty{143}{\giga\byte\per\second}, which gives the ridge of about 71 used in \Cref{tab:roofline-ridge}, and the regime conclusions are identical under either convention.

The archetype arithmetic intensities (fp16) span four orders of magnitude, from 0.25 for elementwise ops to the thousands for conv, and scoring achieved throughput against \texttt{roof@AI} sorts them into three regimes.
Large matmul and conv are the only compute-bound archetypes, so they are the only classes that approach the compute roof.
Convolution gives the ANE's largest measured advantage, about \qty{18.8}{\tera\flop\per\second} and 1370 GFLOP/s/W.
GEMV and decode (AI $\approx 1.0$) lie below every ridge and are memory-bound on both devices, so decode throughput follows whichever device streams weights fastest.
The operator-level GEMV result corroborates the second bandwidth tier but does not bound a fused token loop; \Cref{sec:decode} measures that loop end to end.
Standalone elementwise, norm, and softmax ops reach only a fraction of even the ANE's low bandwidth roof (layer\_norm \qty{18}{\percent}, softmax \qty{63}{\percent}), since its ridge near 424 demands about $3\times$ the AI of the GPU to become compute-bound.

Fusion acts through two separable mechanisms.
One is arithmetic intensity: fusing operators into one compiled program keeps intermediates on-chip and streams weights once, raising effective AI (a standalone conv at about 1076 rises to about 2854 in a fused 3-stack) and moving low-AI points rightward past the ridge, so elementwise, norm and softmax ops run at negligible marginal cost inside a fused program instead of standalone at low utilization.
The other is overhead amortization: a fused graph of any depth lowers to a single \texttt{AneInference} dispatch, so depth amortizes the per-program floor and fixed I/O setup that a standalone dispatch pays in full, while each bridge-induced graph cut becomes a separate dispatch.

The three- and sixteen-conv stacks reach \qty{110}{\percent} and \qty{124}{\percent} of the single-conv saturation rate even though both lie beyond every ridge, where raising AI predicts no further gain.
These values reflect amortized dispatch and I/O cost, not a violated hardware bound.
We therefore call the measured single-dispatch curve an empirical operating envelope, not a hardware upper bound: it is the highest rate one dispatch reaches at that shape, carrying the per-dispatch cost and DRAM I/O that a fused program amortizes, and the fused-chain slope of \Cref{sec:compute-roof} says the same for GEMM.
An entry above \qty{100}{\percent} in \Cref{tab:roofline-place} therefore measures amortization, not a violated bound.

Fusion also produces the gain the roofline does predict from arithmetic intensity.
Fusing a memory-bound neck block (a $1{\times}1$ convolution, group norm, and two gelus) moves its arithmetic intensity from $32.7$--$52.6$ below the weight ridge to $119.9$--$131.2$ above it.
Throughput rises by $3.7\times$ at $C=256$ and $2.3\times$ at $C=512$.

A fused conv or attention graph is efficient on the ANE; the identical ops dispatched standalone are not.

\section{An energy roofline and a predictor}
\label{sec:predictor}

\begin{figure}[t]
  \centering
  \includegraphics{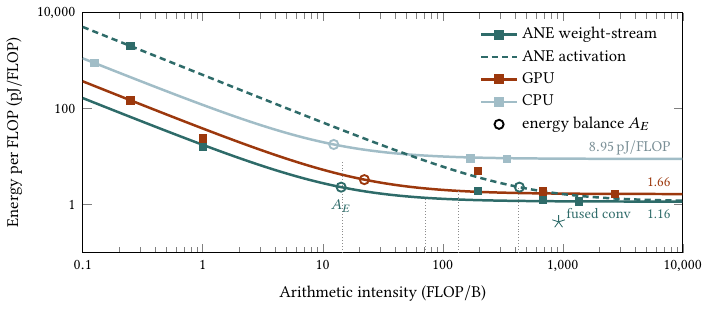}
  \Description{Energy-per-FLOP versus arithmetic intensity for the ANE, GPU, and CPU; the ANE weight-stream path has the lowest energy floor and cheapest byte, the ANE activation path the most expensive byte, and the GPU's energy balance point lies well left of its time ridge.}
  \caption{Energy roofline (energy per FLOP versus arithmetic intensity) on the M5 Pro, fitted from the same active-package-power measurements as the perf/watt tables.
  Solid green is the ANE weight-stream path, dashed green the standalone-activation path; open circles mark each path's energy balance point $\text{AI}_E$ and gray droplines the corresponding time ridge $\text{AI}_T$.}
  \label{fig:energy-roofline}
\end{figure}

The time roofline above bounds throughput; the energy roofline of Choi et al.\ \cite{choi2013energyroofline} bounds energy per operation, $E(\text{AI}) = e_\text{flop} + e_\text{byte}/\text{AI}$, with an energy balance point $\text{AI}_E = e_\text{byte}/e_\text{flop}$ that, like the time ridge, divides a memory-dominated from a compute-dominated regime but in joules, not seconds.
Because every operating point in this paper already carries active package power, the two constants are fitted directly from the ceiling sweeps with no new measurement: $e_\text{flop}$ is the inverse of each device's peak GFLOP/s/W and $e_\text{byte}$ the inverse of its GB/s/W streaming plateau (\Cref{tab:compute-peaks,tab:bw-ceiling}).
\Cref{fig:energy-roofline} draws the result for all three engines, with the ANE split across its two byte paths.

The fitted constants explain the ANE's efficiency: a compute floor of \qty{1.16}{\pico\joule\per\flop} (GPU \qty{1.66}{}) and the cheapest byte in the system, \qty{16.5}{\pico\joule\per\byte} on the weight path (GPU \qty{36.3}{}), while its standalone-activation byte is the most expensive at \qty{503}{\pico\joule\per\byte}, \qty{96}{\percent} of it host dispatch.
The per-watt advantage thus comes from cheaper FLOPs, much cheaper weight bytes, and fusion keeping activation bytes off DRAM, not from the activation path.
The energy balance point lies left of the time ridge on both fast paths ($\text{AI}_E\approx22$ versus $\text{AI}_T\approx134$ on the GPU, $14$ versus $71$ on the ANE weight path), Choi's regime where raising arithmetic intensity yields speed but no further energy savings; only the ANE activation path has the two coincident ($435$, $424$).

The two per-device constants also make the energy roofline predictive.
Given a workload's arithmetic intensity and dispatch path, $E_d(\text{AI}) = e_{\text{flop},d} + e_{\text{byte},d}/\text{AI}$ estimates its energy per FLOP on device $d$, and comparing $E_\text{ANE}$ against $E_\text{GPU}$ identifies the lower-energy engine.
The dispatch path is a second model input and cannot be inferred from arithmetic intensity.
An operator compiled into the fused graph reaches the weight-stream path at \qty{16.5}{\pico\joule\per\byte}; the same operator issued standalone reaches the activation path at \qty{503}{\pico\joule\per\byte}.
The predictor takes that dispatch state as given, and does not derive it.
The device-pair energy crossover is $\text{AI}^\star = (e_{\text{byte,ANE}} - e_{\text{byte,GPU}})/(e_{\text{flop,GPU}} - e_{\text{flop,ANE}})$.
On the ANE weight-stream path the ANE has both the lower compute floor and the cheaper byte, so $\text{AI}^\star$ is negative: the ANE is the lower-energy device at every arithmetic intensity, with no crossover.
The GPU has no separate weight path, so it carries a single streaming byte (\qty{36.3}{\pico\joule\per\byte}); that byte reproduces its measured weight-streaming energy (online supplement, Table~S19), so the crossover compares each device's energy byte, not mismatched paths.
On the standalone-activation path the expensive \qty{503}{\pico\joule\per\byte} moves the crossover up to $\text{AI}^\star \approx 933$, above which only fusing the op back onto the weight path keeps the ANE ahead.
The predictor therefore has one lever: keep FLOP-denominated work on the weight-stream path (compiled and fused), where the ANE is the lower-energy device unconditionally, while a standalone activation-path op below AI 933 belongs on the GPU.

We test the predictor on workloads it was not fitted to, since the constants come only from the ceiling sweeps of \Cref{sec:compute-roof,sec:bandwidth-roof}.
The predictor makes two separable claims, on placement and on magnitude.
On the weight path the crossover is negative, so the model makes one prediction, the ANE, for every weight-streaming workload; the device-map archetypes and the $B=1$ decode loop bear it out (\Cref{tab:device-map,tab:decode}), but this is a single sign, not a test per workload.
The discriminating predictions are on the activation path, where the AI-933 crossover sends a standalone low-intensity op to the GPU.
The standalone group norm (AI 2.5) is GPU-favored in \Cref{tab:device-map}, as predicted, and the below-ridge fusion block switches from the GPU when standalone (AI 32.7 to 52.6) to the ANE when fused (AI 119.9 to 131.2), the measured $3.7\times$ and $2.3\times$ gains confirming the ANE side.

The magnitude claim follows from the roofline being a lower bound on energy per FLOP for single-dispatch work: the predicted joules are accurate near the roof and a one-sided bound elsewhere.
Tested against the held-out device-map classes (online supplement, Table~S19), the predicted energy per FLOP is below measurement on every single-dispatch class, as a lower bound requires.
Near the roof it lands within \qtyrange{7}{28}{\percent} (median \qty{14}{\percent} across both devices, on compute-bound GEMM, long-sequence attention, and the $B=1$ decode loop).
Away from the roof it under-predicts the dispatch-bound rows by large factors, whose package energy is dominated by the fixed per-call cost the roofline omits.
The bound's domain excludes dispatch-amortized graphs: the fused 16-deep conv stack runs at \qty{0.44}{\pico\joule\per\flop}, beneath the fitted floor, because one dispatch covers sixteen layers (\Cref{sec:roofline}).

\paragraph{A dispatch term.} The gap away from the roof is a fixed per-call cost, so we model it.
Writing the roofline per call as $E_\text{call} = e_\text{flop}F + e_\text{byte}B$ with $B = F/\text{AI}$, and adding one constant,
\begin{equation}
E_\text{call} = e_\text{flop}F + e_\text{byte}B + E_\text{disp},
\qquad
T_\text{call} = \max\!\left(\frac{F}{P_\text{peak}}, \frac{B}{BW_\text{peak}}\right) + T_\text{disp},
\label{eq:dispatch}
\end{equation}
where $e_\text{flop}$, $e_\text{byte}$, $P_\text{peak}$ and $BW_\text{peak}$ keep the values already fitted from the ceiling sweeps, so each equation gains exactly one parameter.
Across the held-out classes the work $F$ spans a factor of 1024 while the roofline's residual spans only $2.8$, from \qtyrange{354}{987}{\micro\joule} on the ANE.
The omitted cost therefore does not scale appreciably with work.
The fitted ANE constants are $E_\text{disp} = \qty{681}{\micro\joule}$ and $T_\text{disp} = \qty{136}{\micro\second}$ per call; the GPU values are \qty{1771}{\micro\joule} and \qty{67}{\micro\second}.
Adding them cuts the median absolute error on energy per FLOP from \qty{56.6}{\percent} to \qty{13.7}{\percent}.
Scoring each class with the constant refitted without it gives \qty{16.5}{\percent}.
The dispatch-floor classes carry the improvement: the smallest GEMM moves from \qty{-97.6}{\percent} to \qty{30.1}{\percent} and the single conv from \qty{-97.0}{\percent} to \qty{-8.4}{\percent}, while the near-roof classes stay within \qty{5}{\percent}.

An independent measurement confirms that the fitted $T_\text{disp}$ is a dispatch floor, not a free parameter absorbing error.
The per-program floor of about \qty{70}{\micro\second} (\Cref{sec:decode}) comes from the minimum-latency probe, one dispatch under the most favorable conditions.
\Cref{eq:dispatch} is instead fitted to sustained execution, whose per-call cost is larger by the measured sustained-to-probe ratio of $1.99$ at that shape.
The two agree to \qty{2.5}{\percent}: $70 \times 1.99 = \qty{139}{\micro\second}$ expected against \qty{136}{\micro\second} fitted.
We report the GPU's term as fitted only.
Its independently measured \qty{184}{\micro\second} intercept comes from the stencil step-count sweep, a compiled 32-step program launch, not a single-operator dispatch, and therefore measures a different quantity.

The corrected roofline is thus a placement rule: from a workload's arithmetic intensity and its dispatch path, assign it to the lower-energy engine and fuse to keep it on the weight-stream path.
Two of the three inputs reproduce across generations: the compute floor stays near \qty{1}{\pico\joule\per\flop} (\qty{1.16}{} on the M5 Pro, \qty{0.96}{} on the M4, \qty{1.3}{} on the M2 Pro) and the weight-path placement holds on every measured part (\Cref{sec:three-gen}).
The byte constant does not: the bandwidth-starved base M4 carries a higher effective weight byte (about \qty{34}{} against the M5 Pro's \qty{16.5}{\pico\joule\per\byte}), the byte-bound-versus-element-bound contrast of \Cref{sec:decode}, so the predictor's energy magnitude is host-specific and reported on the fitted M5 Pro alone.

\section{Single-stream device map}
\label{sec:device-map}

From each operator's arithmetic intensity, the corrected roofline predicts the single-stream device map measured here.
The single-stream ($B=1$) device map addresses the first practical choice: for one inference on one input, which engine a given operator class should run on.
\Cref{tab:device-map} reports latency, fp16 relative error, idle-subtracted active package power with CV\%, and perf/watt for the ANE and GPU across thirteen synthetic workload classes; \Cref{tab:real-models} reports three real models with energy per inference.

\begin{table}[htbp]
  \centering
  \footnotesize
  \setlength{\tabcolsep}{1pt}
  \caption{Single-stream device map (M5 Pro, 6 s power window, idle package = 1196 mW); metrics defined in \Cref{sec:methodology}, with CV\% in parentheses in the W columns; fp16 relative error per class is in the online supplement (Table~S20).
Best speed and perf/W per row in \textbf{bold}; GF/W is GFLOP/s per watt, the norm rows elements/s per watt.
Shapes: GEMM $M{\cdot}K{\cdot}N$; attention sequence length $S$ ($D=768$, $H=12$); conv rows are a single $16\to64$ layer and a stack of one $64\to256$ then fifteen $256\to256$ layers (18.4 GFLOP); stencil counts the dense $3\times3$ conv (a strict 5-point count scales every device by $5/9$, leaving ratios unchanged).
$^\dagger$The stencil GPU column is an optimized GPU stencil (shifted-slice arithmetic under \texttt{mx.compile}), not the dense $3\times3$ convolution the ANE runs, and both columns are $N{=}5$ means; the DFT row's single-session ordering is not a verdict (online supplement, formulation-fairness appendix).
$^\ast$Replicated over independent cold starts: the stencil ($N{=}5$), \texttt{group\_norm} and the DFT ($N{=}10$, intervals in \Cref{fig:perfwatt}), and the conv stack, ViT attention, and GEMM-compute rows (online supplement, Table~S17).
The seven unmarked rows are single-session.}
  \label{tab:device-map}
  \begin{tabular}{llrrllll}
    \toprule
    Class & Workload & ANE lat & GPU lat & ANE W (CV) & GPU W (CV) & ANE perf/W & GPU perf/W \\
    \midrule
    GEMM floor & M64$\cdot$K256$\cdot$N256 & \textbf{0.092} & 0.124 & \textbf{2.93 (\qty{11}{\percent})} & 6.54 (\qty{5}{\percent}) & \textbf{15.7} GF/W & 7.8 GF/W \\
    GEMM bandwidth & M128$\cdot$K1024$\cdot$N1024 & \textbf{0.086} & 0.187 & 9.22 (\qty{7}{\percent}) & 9.40 (\qty{6}{\percent}) & \textbf{274} GF/W & 160 GF/W \\
    GEMM compute$^\ast$ & M256$\cdot$K4096$\cdot$N4096 & 1.293 & \textbf{0.461} & \textbf{8.71 (\qty{4}{\percent})} & 33.05 (\qty{8}{\percent}) & \textbf{751} GF/W & 499 GF/W \\
    conv & single $3{\times}3$ ($16\to64$) & \textbf{0.103} & 0.173 & \textbf{5.99 (\qty{12}{\percent})} & 7.61 (\qty{3}{\percent}) & \textbf{24.5} GF/W & 14.7 GF/W \\
    conv$^\ast$ & ResNet $3{\times}3{\times}16$ ($64\to256$) & \textbf{0.757} & 3.159 & \textbf{10.12 (\qty{6}{\percent})} & 32.70 (\qty{2}{\percent}) & \textbf{2289} GF/W & 175 GF/W \\
    attention$^\ast$ & ViT block $S=197$ & \textbf{0.230} & 0.269 & \textbf{8.50 (\qty{17}{\percent})} & 16.01 (\qty{11}{\percent}) & \textbf{522} GF/W & 197 GF/W \\
    attention & long-seq $S=512$ & 0.817 & \textbf{0.390} & \textbf{6.89 (\qty{7}{\percent})} & 23.22 (\qty{11}{\percent}) & \textbf{561} GF/W & 305 GF/W \\
    norm & layer\_norm (197,768) & 0.195 & \textbf{0.176} & \textbf{5.76 (\qty{12}{\percent})} & 9.03 (\qty{6}{\percent}) & \textbf{1.25e8}/W & 9.24e7/W \\
    norm & rms\_norm (197,768) & 0.373 & \textbf{0.151} & \textbf{3.44 (\qty{6}{\percent})} & 7.75 (\qty{6}{\percent}) & \textbf{1.13e8}/W & 1.12e8/W \\
    norm$^\ast$ & group\_norm (1,256,$64^2$) & 0.350 & \textbf{0.196} & \textbf{7.89 (\qty{7}{\percent})} & 10.82 (\qty{9}{\percent}) & 3.63e8/W & \textbf{3.85e8}/W \\
    sci$^\ast$ & 5-pt stencil $256^2$, 32 steps$^\dagger$ & \textbf{0.137} & 0.382 & \textbf{7.54 (\qty{14}{\percent})} & 11.58 (\qty{1}{\percent}) & \textbf{37.1} GF/W & 7.86 GF/W \\
    sci & Jacobi $n{=}512$, 25 iters & \textbf{0.170} & 0.285 & \textbf{5.24 (\qty{13}{\percent})} & 9.98 (\qty{6}{\percent}) & \textbf{14.5} GF/W & 3.98 GF/W \\
    sci$^\ast$ & DFT-as-matmul $N{=}1024$ & \textbf{0.176} & 0.185 & \textbf{4.06 (\qty{25}{\percent})} & 8.24 (\qty{5}{\percent}) & 2.45 GF/W & \textbf{2.99} GF/W \\
    \bottomrule
  \end{tabular}
\end{table}

\begin{table}[htbp]
  \centering
  \footnotesize
  \setlength{\tabcolsep}{2.2pt}
  \caption{Real models (latency ms; energy mJ/inference; all GPU arms fp16 unless marked).
  \texttt{MPS} is torch-MPS (\texttt{.half()}) and \texttt{MLX-c} is graph-compiled MLX (\texttt{mx.compile}); MiniLM and ViT-B/16 run as MLX ports loading the same weights (online supplement, formulation-fairness appendix).
  The ANE and \texttt{MPS} energies are means over $N{=}16$ cold starts, and \emph{adv.\ MPS} is their geometric-mean ratio with a \qty{95}{\percent} interval, whose width is set mostly by the GPU arm (online supplement, Table~S6).
  The fp32 \texttt{MPS} column$^\dagger$ is single-session and carries no interval.
  The \texttt{MLX-c} energies and \emph{adv.\ MLX-c} are paired repetitions interleaved with the ANE within one session ($N{=}5$ for ResNet-18, $N{=}10$ for the others), with a \qty{95}{\percent} $t$-interval on the mean ratio.
  Each \emph{adv.\ MLX-c} ratio pairs a compiled-MLX run with an ANE run from the same session, so it is not the quotient of the printed cold-start ANE energy (the in-session ANE energies are in the online supplement, formulation-fairness appendix).
  Latency is the device-map probe, except \texttt{MLX-c}, which is the sustained time of its energy window.}
  \label{tab:real-models}
  \begin{tabular}{lrrrlrrrrll}
    \toprule
     & \multicolumn{3}{c}{Latency (ms)} & & \multicolumn{4}{c}{Energy (mJ/inf)} & \multicolumn{2}{c}{ANE energy advantage} \\
    \cmidrule(lr){2-4}\cmidrule(lr){6-9}\cmidrule(lr){10-11}
    Model & ANE & MPS & MLX-c & ANE relerr & ANE & MPS fp32$^\dagger$ & MPS & MLX-c & adv.\ MPS & adv.\ MLX-c \\
    \midrule
    ResNet-18 & \textbf{0.330} & 2.033 & 1.645 & 2.3e-3 & \textbf{2.30} & 39.1 & 21.3 & 32.6 & $8.9\times$ $[7.5, 10.7]$ & $14.6\times$ $[12.1, 17.2]$ \\
    MiniLM (1 sent) & \textbf{0.525} & 1.920 & 0.650 & 5.5e-3 & \textbf{2.74} & 28.8 & 17.6 & 11.5 & $6.7\times$ $[5.9, 7.6]$ & $3.4\times$ $[3.0, 3.9]$ \\
    ViT-B/16 & 18.34 & 15.92 & \textbf{2.959} & 7.8e-3 & \textbf{65.7} & 1382.7 & 577.9 & 100.4 & $9.4\times$ $[7.9, 11.1]$ & $1.5\times$ $[1.3, 1.8]$ \\
    \bottomrule
  \end{tabular}
\end{table}

\begin{figure}[t]
  \centering
  \includegraphics{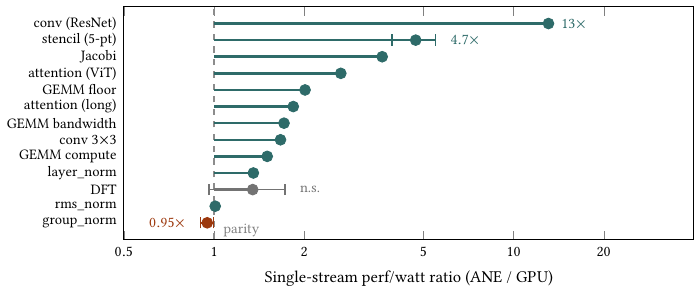}
  \Description{Sorted lollipop chart of the ANE-over-GPU single-stream performance-per-watt ratio for thirteen workload classes against a parity line; the ANE is more efficient on 11 of 13 classes, by up to 13x on the ResNet-like conv stack, the GPU is ahead on group norm at 0.95x, and the DFT is statistically indeterminate with a confidence interval spanning parity.}
  \caption{Single-stream perf/watt advantage, ANE divided by GPU, per workload class on the M5 Pro (\Cref{tab:device-map}), sorted and drawn against a parity line.
  The ANE is more energy-efficient on 11 of 13 classes, by up to $13\times$ on the ResNet-like convolution stack; the stencil is scored against an optimized GPU stencil, which places it at $4.7\times$.
  Error bars are 95\% CIs over independent cold starts for the stencil, the DFT, and \texttt{group\_norm}; the DFT interval spans parity, so that class supports no directional verdict (n.s.).
  The conv stack, ViT attention, and GEMM-compute classes are also replicated (perf/watt differences with CIs in online supplement Table~S17), and \Cref{tab:device-map} marks every replicated class.}
  \label{fig:perfwatt}
\end{figure}

The single-stream map shows that the ANE draws less power on every in-graph-expressible workload tested (\Cref{fig:perfwatt}), a $1.02$ to $3.8\times$ reduction in active package power, even on the workloads where the GPU is faster.
On large square fp16 GEMM ($K=4096$) the GPU is $2.8\times$ faster and more accurate (\num{3.6e-4} versus the ANE's \num{2.7e-2}), yet the ANE still does the work on \qty{8.71}{\watt} against the GPU's \qty{33.05}{\watt}.
The ANE has the lower latency on convolution, convolution stacks ($4.2\times$ on the ResNet-like block), the stencil ($2.8\times$ against an optimized GPU stencil), short attention, mid-size GEMM, and the smaller real models (ResNet-18 $5.0\times$ and MiniLM $1.2\times$ against compiled fp16 MLX).
It has the higher perf/watt on 11 of 13 classes, by the largest margin on the ResNet-like conv stack ($13\times$), then the stencil ($4.7\times$) and the Jacobi solve ($3.6\times$).
The remaining two classes sit at or near parity, and repetition separates them.
Over ten independent cold starts the GPU advantage on \texttt{group\_norm} is small and reproducible ($0.949$, 95\% CI $[0.900, 0.997]$).
The DFT ratio is $1.35$ with a CI of $[0.96, 1.73]$ that spans parity, so that class carries no directional verdict (online supplement, formulation-fairness appendix).
The exception among the real models is ViT-B/16, on which the GPU is faster (\Cref{tab:real-models}).
Against torch-MPS fp16 the margin is small ($15.9$ versus the ANE's \qty{18.3}{\milli\second}) and the ANE's energy advantage is $9.4\times$.
Graph-compiled MLX runs the model in \qty{2.96}{\milli\second}, $6.2\times$ faster than the ANE, and the ANE's energy advantage drops to $1.5\times$ (95\% CI $[1.3, 1.8]$).
Compiled MLX also halves MiniLM's advantage, from $6.7\times$ to $3.4\times$, though the ANE remains the faster device there.
The GPU has the lower latency on large square GEMM, long-sequence attention, and the small norm reductions, and is more accurate in fp16 everywhere.

A precision-versus-speed rule follows from the measured fp16 accuracy.
Use the ANE's fp16 speed and efficiency where the math is well-conditioned and not a large square GEMM (convolution, stencils, short attention, mid-size GEMM, encoders); there its relative error of \num{1e-3} to \num{1e-2} is the same order as GPU fp16 and it is usually both faster and lower-power.
Hand large compute-bound square matmuls ($K \geq 4096$) and long sequences to the GPU.
Use fp32 wherever cancellation can dominate (an adversarial, ill-conditioned reduction), where fp16 error is unbounded.
The headline perf/watt is the conservative total-package number that charges the ANE for its own host CPU dispatch, and the ANE is the more efficient device on it for 11 of the 13 device-map classes, so the efficiency is silicon; the ANE-rail-only silicon bound is $1.2$ to $4\times$ larger (online supplement, dual-bound table).

\section{Cross-generation reproduction}
\label{sec:three-gen}

The single-stream watt harness also runs on three further hosts: an M2 Pro (h14 ANE, a 16~GB Mac mini), an M1 Max (h13, a MacBook Pro), and an M4 (h16, a second 16~GB Mac mini, base part), so the reproduction spans four generations and both chassis classes.
The online supplement gives the per-class maps (Tables~S11--S13); h16's three real-model rows were not obtained, as ResNet-18 hung at compile on that part.
Those maps cover the eleven synthetic classes common to all four hosts, two fewer than \Cref{tab:device-map}, since Jacobi and the DFT-as-matmul were added to the reference map afterwards and are M5 Pro only.

Every substantial-workload result reproduces, where substantial means the classes above the dispatch floor: conv stacks, attention, compute-scale GEMM, the stencil, and the real models where they ran.
Every substantial class favors the same device on all four generations, and the ANE is the more efficient device in each case.
The conv-stack perf/watt ratio stays between $12.9$ and $14.5\times$ across the four parts.
Non-reproducing rows are confined to the dispatch-floor and tiny-norm classes, where host dispatch dominates package power.
These shift toward the GPU on the older parts and back on the newest.
Across the eleven classes, the faster device matches the M5 map on 7 and the more efficient device on 7, with 3 matching on both axes.
On h16 the disagreements shrink instead of growing, the GPU keeping a perf/watt edge only on \texttt{group\_norm}.

The bandwidth, serving, and decode sweeps reproduce in structure on the desktop parts.
Both h14 and h16 show the same low-activation-ceiling asymmetry.
On h14 the batched-serving throughput crossover falls near $B=36$, with no throughput-per-watt crossover on any of the four workloads.
On h16 the ANE leads in tokens-per-watt at every batch, by a wider margin than on the M5 Pro.
It does not run on h14, whose \num{16384} maximum tensor dimension the \num{32000}-token vocabulary head exceeds.
A fused-GPU baseline (\texttt{mx.compile}) shows the large gaps are not MLX launch overhead: fusion helps only the small elementwise rows.

Formulation is a separate confound, and one class is bound by it: expressing the 5-point stencil as a dense $3\times3$ convolution costs the GPU $3.3\times$ in time and $2.9\times$ in power, moving that row from $43.8\times$ to $4.7\times$ (online supplement, fused-GPU and formulation-fairness appendices).

\section{Batched serving}
\label{sec:serving}

The device map of \Cref{sec:device-map} is the single-stream ($B=1$) choice.
Serving is different: a deployment batches requests.
We sweep batch size $B \in \{1, 4, 16, 64, 256\}$ with true batching, one compiled program with a real batch dimension, not multi-stream, across four serving workloads, and report the two deployment-relevant crossovers: the batch at which GPU throughput first exceeds the ANE's, and the batch at which GPU throughput-per-watt does.
Both devices run the same fp16 math at the same $B$, warm, steady-state; accuracy is flat in $B$.

The throughput crossover appears on three of the four workloads, and the throughput-per-watt crossover on none of them within the swept range.
The online supplement (Tables~S7--S10) gives the sweeps; \Cref{tab:crossover} summarizes the crossovers.

\begin{table}[htbp]
  \centering
  \small
  \caption{Crossover summary: the batch at which GPU throughput, and GPU throughput-per-watt, first exceed the ANE's, per serving workload.
A \textbf{never} entry marks a workload where the ANE leads at every swept $B$, with the ANE/GPU multiplier range in parentheses.}
  \label{tab:crossover}
  \footnotesize
  \setlength{\tabcolsep}{6pt}
  \begin{tabular}{lll}
    \toprule
    Workload & Throughput crossover $B^*$ & Throughput/W crossover \\
    \midrule
    Vision (conv classifier) & \textbf{never} ($3.1$--$5.7\times$) & \textbf{never} ($6$--$10\times$) \\
    Encoder (transformer block) & \textbf{$\sim$23--31} (replicate $31.1\pm1.3$) & \textbf{never} ($1.9$--$5.4\times$) \\
    Attention (self-attention block) & \textbf{small batch} (replicate $2.23\pm0.02$) & \textbf{never} ($1.7$--$2.4\times$) \\
    Batched GEMM & \textbf{$\sim$5} ($B=4\to16$) & \textbf{never} ($1.06\times$ at $B=64$; \Cref{sec:replicate}) \\
    \bottomrule
  \end{tabular}
\end{table}

For vision both crossovers are absent: the ANE leads at every batch on throughput and perf/watt.
For the transformer workloads the GPU's raw throughput crosses above the ANE's at small-to-moderate batch (attention near $B=2$, encoder near $B=31$; \Cref{sec:replicate}), while the ANE keeps the higher throughput-per-watt across the full swept range, and on bare large-batch GEMM it retains about \qty{6}{\percent} (resolved under the replicate protocol of \Cref{sec:replicate}).
The GPU is the throughput device for batched transformers; the ANE's serving role there is the low-batch and energy-capped regime.
The throughput crossover depends on the workload, falling near $B=2$ for a self-attention block and near $B=31$ for a transformer encoder.
Above its own crossover the GPU maximizes throughput.
The ANE remains preferable whenever sequences-per-joule is the target, and for vision under both metrics.

\section{End-to-end LLM decode}
\label{sec:decode}

The serving sweeps above measure encoder, attention, and conv blocks.
The remaining serving case, and the one where a single operating point most easily misleads, is autoregressive LLM decode, the per-token forward pass of a generative transformer.
The synthetic GEMV operating point (\Cref{tab:roofline-place}: ANE 112 $>$ CPU 81 $>$ GPU 75 GFLOP/s at AI $\approx 1.0$) suggests an ANE single-stream decode advantage.
To test whether that op-level point transfers, we measured a full decode loop: one token is a complete 8-layer transformer forward at sequence position 1, with the key--value (KV) cache (length 256) fed as graph inputs.
The configuration is a TinyLlama/GPT-small class model (\texttt{d\_model=2048}, 8 layers, 16 heads (\texttt{d\_head=128}), a $4\times$ feed-forward network (FFN, \texttt{d\_ff=8192}), vocab 32000, KV-len 256) at 1.22 GFLOP/token ($B=1$) and 39.1 GFLOP/token ($B=32$).
The ANE attention is built from \texttt{bmm} and \texttt{softmax} primitives, not \texttt{af.sdpa} (which needs equal-shape query/key/value tensors and so does not apply to the $M=1$ decode step).
The GPU runs MLX fp16, the CPU numpy/Accelerate fp32 (the reference); ANE and GPU run the same fp16 math.
\Cref{tab:decode} reports the result with the same per-engine energy accounting as the rest of the paper.

\begin{table}[htbp]
  \centering
  \caption{End-to-end LLM decode, batch sweep (M5 Pro; 8-layer forward per token at seq pos 1; KV-len 256; ANE attention = bmm+softmax; idle pkg $\approx$ 1480 mW).
  Throughput (tok/s) and idle-subtracted total-package efficiency (tok/s/W) for the ANE (fp16 and per-channel int8 weight-stream) and GPU (MLX fp16).
  The fp32 relative-error reference is numpy/Accelerate.
  Best throughput and best efficiency per batch in \textbf{bold}.
  fp16-versus-fp32 relative error $\approx$ \num{2.5e-2} (ANE fp16), \num{5.0e-2} (ANE int8), \num{3.2e-3} (GPU).
  The GPU's $B=2$ efficiency dip is a power-state step (the GPU rail pins at \qty{19.9}{\watt} for all $B\geq2$ before batching amortizes it).}
  \label{tab:decode}
  \begin{tabular}{rrrrrrr}
    \toprule
    B & ANE tok/s & int8 tok/s & GPU tok/s & ANE-fp16 tok/s/W & ANE-int8 tok/s/W & GPU tok/s/W \\
    \midrule
    1 & 115 & 123 & \textbf{200} & 35.9 & \textbf{44.7} & 19.3 \\
    2 & 204 & 204 & \textbf{231} & 50.6 & \textbf{68.7} & 9.9 \\
    4 & 341 & 342 & \textbf{431} & \textbf{68.6} & 64.8 & 18.5 \\
    8 & 517 & 515 & \textbf{834} & 88.2 & \textbf{100.3} & 36.9 \\
    16 & 704 & 698 & \textbf{1444} & 106.2 & \textbf{113.7} & 65.1 \\
    32 & 847 & 850 & \textbf{2322} & 88.0 & 105.4 & \textbf{116.4} \\
    \bottomrule
  \end{tabular}
\end{table}

The op-level GEMV ordering does not carry to end-to-end decode throughput.
Measured on a real loop, the GPU has the higher throughput at every batch in fp16 and int8, $1.74\times$ the ANE at $B=1$ (200 versus 115 tok/s), widening to $2.7\times$ at $B=32$ (2322 versus 847), the reverse of the single-GEMV microbenchmark.

The per-token time budget explains both the deficit and the reversal.
The decode forward compiles to a single fused program per token step, so the per-program dispatch floor~\cite{ane_guide} (about \qty{70}{\micro\second} on this part) is paid once per token, under one percent of the \qty{8.7}{\milli\second} fp16 token.
The budget is instead dominated by the weight stream: the model as measured carries 602.4\,M weight parameters across the eight decoder blocks and the vocabulary projection, \qty{1.20}{\giga\byte} in fp16.
The batch sweep separates the costs cleanly, since per-step latency is linear in batch with intercept \qty{7.9}{\milli\second} and slope \qty{0.93}{\milli\second} per stream (residuals under \qty{1.5}{\percent} at all six batch points).
The intercept is the batch-invariant weight stream, \qty{1.20}{\giga\byte} at an effective \qty{154}{\giga\byte\per\second}.
This agrees with the standalone GEMV sweep after removing 56 amortized dispatch floors from the \qty{7.6}{\milli\second} total for its 57 constituent GEMVs.
The slope is the per-stream traffic, chiefly the \qty{16.8}{\mega\byte} KV cache fed through the slower activation path.
Single-stream decode on the ANE is therefore weight-stream-bound; dispatch would dominate only if the stack were issued one program per operation, costing 57 floors and capping the loop near 80 tok/s.

The GPU streams the same way faster: at 200 tok/s it moves the same \qty{1.20}{\giga\byte} in \qty{5.0}{\milli\second}, about \qty{240}{\giga\byte\per\second}, near its achieved streaming ceiling.
Decode is weight-stream-bound on both devices, as the roofline predicts, and the GPU's higher decode throughput follows from its faster weight stream.
The GEMV microbenchmark corroborates the weight-stream tier (\Cref{sec:weight-stream,sec:roofline}), though standalone small-$K$ GEMVs are dispatch-floor-dominated and do not bound the fused loop.

int8 weight streaming identifies which resource binds the stream.
Decode is the AI $\approx 1.0$ memory-bound regime, so per-channel int8 (half the weight bytes) is the one change that could move throughput, not energy alone.
It barely does: int8 lifts $B=1$ throughput by \qty{7}{\percent} (115 $\to$ 123 tok/s) and then matches fp16 for every $B \geq 2$ (within \qty{1}{\percent}).
A purely byte-bandwidth-bound stream would roughly double under this compression, and the measured change is near null.
Combined with the weight-stream-bound budget above, that places the limit at the rate weight elements pass the dequantize-and-multiply stage, about 77\,Gelem/s, and not at DRAM bytes.

That ceiling belongs to the dequantize stage, not to the weight stream, and a third encoding separates the two.
Streaming the same weights as an int4 lookup table quarters the weight blob (\qtyrange{1149}{287}{\mega\byte}) and raises $B=1$ throughput to 203 tok/s, $1.65\times$ the int8 rate, where an element-rate limit predicts no change.
That is level with the GPU's fp16 rate of 200 tok/s; we did not measure an int4 GPU arm, so this is not a like-for-like throughput comparison.
The encodings compile to different constant operators, \texttt{constexpr\_affine\_dequantize} for int8 and \texttt{constexpr\_lut\_to\_dense} for int4, and the measured rates separate with them: fp16 streams at \qty{140}{\giga\byte\per\second} and 70\,Gelem/s, int8 at \qty{74}{\giga\byte\per\second} and 74\,Gelem/s (the 77\,Gelem/s above, re-measured), and int4 at \qty{61}{\giga\byte\per\second} and 122\,Gelem/s.
The fp16 stream is thus byte-bound at the \qtyrange{130}{143}{\giga\byte\per\second} GEMV plateau, the int8 stream is element-rate-bound below it, and the lookup-table stream clears both.
The ceiling is therefore a property of the weight encoding, not of precision: which constant operator the encoding lowers to sets the rate, and the three operators measured give three rates.
The int4 program is inaccurate on this synthetic model, disagreeing with the fp64 reference on the greedy token at a rank correlation of $0.82$ against $0.998$ for int8, so its output alone does not establish that the encoding streams every weight.
ResNet-18 does.
Requested at the default budget the encoding falls back to int8 and reproduces the int8 row exactly (\qty{69.65}{\percent} top-1, \qty{2.33}{\percent} label disagreement); engaged, it is a distinct and working classifier at \qty{67.21}{\percent} top-1 over all \num{50000} validation images (\Cref{sec:methodology}).
A path that skipped weights would not classify at that rate.
The synthetic model's error reflects randomly initialized weights, which give a 4-bit palette no cluster structure to exploit.
The online supplement (Table~S23) reports the four encodings with weight blobs read from the compiled artifact.
This measurement is one configuration at $B=1$ on the M5 Pro and is not checked across generations, which matters because int8 decode is already at a different bound on the base M4.

What compression yields instead is energy savings: halving the weight bytes cuts energy per token by \qty{20}{\percent} at $B=1$ (27.8 $\to$ \qty{22.4}{\milli\joule}, about \qty{9}{\pico\joule} per avoided byte, an LPDDR-scale figure) and improves the ANE's tok/s/W at most batches (by up to \qty{36}{\percent}, though it dips about \qty{6}{\percent} at $B=4$), at roughly $2\times$ the fp16 relative error (about \num{5e-2}).
int8 does not change the GPU's throughput advantage at any batch.
This element-rate ceiling is host-specific: on the more bandwidth-starved base M4 (h16) int8 does raise ANE decode throughput ($1.3$ to $2\times$), which shows that part's decode is at the DRAM-byte bound, not at this element-rate limit (\Cref{sec:three-gen}).
The accuracy cost is bounded: int8 holds the greedy token \qty{92}{\percent} of the time versus the fp64 reference.

The energy axis is a measured crossover, not an immediate switch.
The ANE has the higher tok/s/W at every batch up to $B=16$ in both precisions, int8 by $2.3\times$ at $B=1$ (44.7 versus 19.3), $2.7\times$ at $B=8$ (100.3 versus 36.9), and still $1.75\times$ at $B=16$ (113.7 versus 65.1), with fp16 close behind (35.9 at $B=1$, 106.2 at $B=16$).
The GPU's efficiency crosses above only at $B=32$ (116.4 versus 105.4 int8, 88.0 fp16), where the ANE's package draw grows with the host CPU feeding $B \times \qty{16.8}{\mega\byte}$ of KV cache per step.
(The GPU's dip at $B=2$ is a power-state step: its rail jumps from \qty{6.3}{} to a pinned \qty{19.9}{\watt} before batching amortizes the full-power clocks.)
Sweeping the KV cache from 128 to 2048 positions at $B=1$, per-step latency is linear in context on both devices, and the fit depends on where the cache lives.
Holding it resident in the engine's input buffers, the ANE runs at \qty{7.75}{\milli\second} plus \qty{1.34}{\micro\second} per position against the GPU's \qty{4.74}{\milli\second} plus \qty{0.73}{\micro\second}.
Both intercepts recover the batch-invariant weight stream on an axis it was not fitted against, and the KV term reaches the weight term only at $s_{kv}\approx5800$ on the ANE and $\approx6500$ on the GPU, so single-stream decode is weight-stream-bound across any context either part will hold.
The ANE's KV path is $1.8\times$ slower per position, the activation path of \Cref{sec:bandwidth-roof} and not the weight path, and its efficiency advantage is \numrange{2.6}{2.0} across the sweep, with no crossover.

Re-feeding the cache to the engine on every token instead, as a graph input instead of a resident buffer, is what the roofline predicts it to be and is worth separating from the silicon.
The cost is \qty{0.51}{\milli\second} and \qty{0.92}{\watt} at $s_{kv}=256$, rising to \qty{4.38}{\milli\second} and \qty{2.18}{\watt} at 2048, where the transfer is \qty{128}{\mega\byte} per token.
It also triples the fitted per-position slope, to \qty{3.49}{\micro\second}, and it erases the ANE's efficiency advantage by $s_{kv}\approx1300$.
The signature is a falling engine rail, \qtyrange{1878}{1428}{\milli\watt}, beside a rising package, \qtyrange{3.75}{5.86}{\watt}: the engine idles while the host marshals.
Held resident, the rail is flat at about \qty{1.93}{\watt} and the package falls to \qty{3.68}{\watt} at the same point.
This host-feeding term also produces the batch crossover above; it is an implementation cost, not a property of the part.

Batched decode gives the GPU the throughput axis immediately but the energy axis only past $B\approx16$.
At small batch, on-device decode uses 2 to $3\times$ less energy on the ANE.
That advantage ends near $B\approx16$ on the batch axis, and across context length it persists only while the caller keeps the cache resident.

These magnitudes are for the one model measured.
Decode here is weight-stream-bound because the \qty{1.20}{\giga\byte} weight stream is about $70\times$ the per-token KV traffic at $s_{kv}=256$, and the throughput ordering follows the two devices' weight-stream rates.
A model whose weight bytes dominate its KV bytes should order the same way.
The energy crossover depends on the ratio of per-stream KV and host-feeding cost to the weight stream, which changes with width, depth, attention layout, and vocabulary (the \num{32000}-token vocabulary projection alone is \qty{65.5}{M} of the \qty{602.4}{M} streamed parameters), and we make no claim for where it falls on other models.

\section{Discussion}
\label{sec:discussion}

The corrected roofline has a corollary the four sweeps confirm: the ANE's performance-per-watt advantage is regime-dependent, set by where a workload's arithmetic intensity falls relative to the two ridges.
It is largest in the compute-bound, low-to-moderate-utilization regime, where the ANE is both faster and more efficient (\Cref{tab:device-map}, and ResNet-18 and MiniLM in \Cref{tab:real-models}).
Four regimes favor the GPU.
It has the higher raw throughput once a device is saturated, where it also takes a small conv perf/watt edge (\Cref{sec:compute-roof}), and it leads the standalone memory-bound regime, where the ANE's activation path runs about $10\times$ below the GPU's unified-memory rate (\Cref{sec:bandwidth-roof}).
It also leads batched-transformer serving (\Cref{sec:serving}) and, in the measured decoder, decode throughput at every batch (\Cref{sec:decode}).
Vision serving is the exception.
ViT-B/16 and single-stream decode are energy-only exceptions: the ANE uses less energy but has the higher latency.
The decode result holds across the measured context range while the KV cache stays resident.
The two-ridge structure accounts for this split, and HeteroLLM \cite{chen2025heterollm} reports the same split independently on a different NPU (\Cref{sec:related}).

This corrects the community guidance \cite{hollemans_neuralengine,apple2022anetransformers}.
The 38-TOPS shorthand (an M4 figure) reduces a regime-dependent engine to one peak number.
The measurements also contradict the guidance that the ANE is worthwhile mainly for big models: its largest advantages are in low-batch, latency- and energy-capped on-device inference.
The claim that the ANE is more efficient is true in the compute-bound regime, narrows to a small GPU edge on conv at peak, and is false in the standalone memory-bound regime.

Fusion is the program-level control on arithmetic intensity, and so on regime: fusing operators into one compiled program raises effective AI and moves the operating point past the ridge (\Cref{sec:roofline}).
An arithmetic-intensity-aware scheduler should place work by arithmetic intensity and target utilization.
Operators that would otherwise be memory-bound on the ANE should be fused into a larger graph where possible.
The choice is also constraint-dependent.
Under a tight power or energy cap the ANE is preferable for every serving workload measured here, though on bare large-batch GEMM by only about \qty{6}{\percent}.
Without such a cap the GPU is preferable past each workload's crossover.
CoreML exposes model-level placement through \texttt{computeUnits}, which can assign vision, convolution, and small-encoder models to the ANE.
It does not expose the per-operator fusion control that moves a memory-bound operator across the ridge, nor does it report the per-engine energy measured here.

\paragraph{Replication.}\label{sec:replicate}
Re-measured over $N=5$ independent cold-start runs, conv at peak resolves to a \qty{5}{\percent} GPU edge and batched GEMM at $B=64$ to a \qty{6}{\percent} ANE edge (both Holm-significant), the large ANE margins (conv stack, attention, GEMM-compute) confirm, and both transformer crossovers are stable ($B^*=2.23\pm0.02$ and $31.1\pm1.3$).
The full protocol and per-claim table are in the online supplement (Table~S17).

\section{Related work}
\label{sec:related}

The analysis descends from the roofline model \cite{williams2009roofline}, extended to energy by Choi et al.\ \cite{choi2013energyroofline} and to cache-aware, hierarchical, and instruction-level variants \cite{ilic2014carm,hierarchical_roofline,ding2019instructionroofline}, with an efficiency-oriented framing for ML accelerators \cite{verhelst2025rooflines}.
The ANE exposes two effective bandwidths, so the single-line form does not describe it; the multi-tier roofline \cite{hierarchical_roofline} is what the data re-derive for this NPU (\Cref{sec:roofline}).
We build the roofs empirically from microbenchmarks, not datasheet peaks \cite{lo2014ert}, and the weight-path-versus-activation-path split is the energy-per-data-movement dataflow taxonomy \cite{chen2016eyeriss} seen from outside the black box.
Pagoda \cite{prashanthi2025pagoda} builds dual time-and-energy rooflines on Jetson, and RooflineBench \cite{bi2026rooflinebench} scores on-device LLMs by operational intensity but is throughput-only and does not isolate the ANE.

Smartphone-AI benchmarking was popularized by AI-Benchmark \cite{ignatov2019aibenchmark} and standardized, with a power methodology, by MLPerf Inference \cite{reddi2020mlperf}, though that line reaches the ANE only through CoreML and cannot isolate it.
Seshadri et al.\ \cite{seshadri2022edgetpu} report throughput, utilization, and energy for Google's Edge TPU.
Edge-AI rooflines have targeted Jetson and Coral \cite{edgeai2024roofline}, mobile-NPU LLM inference has been studied at the systems level and in GPU-NPU hybrids \cite{xu2025npullm,hybe2025}, and sustained-load mobile energy has been measured across NPU and GPU including Apple devices \cite{tummalapalli2026edge,laskaridis2024melting}, none isolating the ANE rail.

HeteroLLM \cite{chen2025heterollm} characterizes a Qualcomm Hexagon NPU with an Adreno GPU and finds the same regime split with absolute power, compute-bound prefill to the NPU and memory-bound decode to the GPU; that an independent study reaches this on a different NPU is consistent with the device map reflecting a general property of NPU-versus-GPU economics, not of one vendor.
For the energy methodology, idle-subtracted net energy has precedent in NPU-MCU work \cite{fanariotis2025npumcu}, and programmatic energy measurement on Apple silicon has been discussed but not resolved into a validated estimator \cite{mlenergy_macprofiling}.

A small body of concurrent work establishes direct or reverse-engineered ANE access: Orion \cite{kumaresan2026orion}, the maderix M4-ANE writeups \cite{singh2026anem4}, and the Asahi-Linux ANE driver \cite{yoon_ane_asahi}; neither direct access nor the observation that the engine can be energy-efficient is new here.
Orion reports ANE throughput for LLM training and inference, with an ANE-versus-CPU speed comparison only and no power or roofline measurements.
It independently finds CPU decode outpacing ANE decode on an M4 Max (283 versus 170 tok/s).
Our decode loop compares the ANE against the GPU, not the CPU, but finds the same direction, a faster companion engine, on a different chip and access path (\Cref{sec:decode}), and attributes the deficit to the ANE's weight-stream rate, not to \texttt{aneforge}.
NPUMoE \cite{benazir2026npumoe} runs mixture-of-experts on the ANE through CoreML, as do Apple's transformer guidance \cite{apple2022anetransformers} and the community knowledge base \cite{hollemans_neuralengine}.
Pagoda \cite{prashanthi2025pagoda} and HeteroLLM \cite{chen2025heterollm} are the closest prior characterizations.
Neither isolates the ANE rail in a per-engine roofline with idle-subtracted, package-level energy across the single-stream, batched, compute-saturation, and bandwidth regimes.

\section{Conclusion}
\label{sec:conclusion}

We presented a per-engine energy roofline characterization of the Apple Neural Engine against the Apple GPU and CPU on an M5 Pro, with the single-stream watt map reproduced on an M1 Max, an M2 Pro, and an M4.

The ANE's efficiency advantage is regime-dependent.
It is faster and lower-power for compute-bound work at low-to-moderate utilization, including conv, attention, mid-size GEMM, ResNet-18, and the MiniLM encoder; on ViT-B/16 it is only the lower-energy device, by $1.5\times$ against compiled MLX.
The GPU has the higher throughput at saturation and at transformer-serving scale, a small perf/watt edge on conv at peak, and both higher speed and higher efficiency for standalone memory-bound operators.
The ANE, as reachable today, has two effective bandwidths (about \qty{24}{\giga\byte\per\second} standalone-activation, \qtyrange{112}{143}{\giga\byte\per\second} weight-stream) and two corresponding ridges, the fast one below the GPU's.
Given arithmetic intensity and dispatch path, the energy roofline predicts the lower-energy engine and bounds its energy from below.
Its median error is \qty{14}{\percent} near the roof, and it stays conservative elsewhere within its stated domain.
ML-systems schedulers should place work by arithmetic intensity and utilization, not by operator name.

\subsection*{Limitations of this work}
\label{sec:limitations}

The saturation, bandwidth, serving, and decode magnitudes are from the one Apple M5 Pro (h17s); their structure reproduces across the four generations (\Cref{sec:three-gen}), but the absolute multipliers are host- and thermal-dependent and should not be quoted for other parts, and the cross-generation reproductions are single- or dual-run point estimates, and are not replicated under the cold-start protocol.
The end-to-end decode results are for one model configuration (one size, one \num{32000}-token vocabulary, synthetic weights), and their magnitudes and energy crossover are not claimed for other models (\Cref{sec:decode}).
The activation-path energy crossover (AI$\approx933$) is computed from the two fitted constants and lies well beyond the measured range, so it is an extrapolation, not a measurement.
Both power estimators are modeled SoC telemetry, not an external meter, though they cross-validate (\Cref{sec:methodology}, online supplement Table~S1), and the ANE bandwidth ceiling is a path limit of \texttt{aneforge}'s single-program streaming, not a claim about the silicon DMA.
This is a plugged-in, actively cooled host exposing only an OS thermal-pressure signal, so a fanless or battery-constrained device may behave differently.
The access path uses undocumented but unprotected user-space runtime interfaces~\cite{ane_guide}.
This work characterizes performance only and crosses no content-protection or security boundary.
Despite these limitations, the contested small-margin verdicts are settled under the cold-start replicate protocol with formal inference attached (\Cref{sec:replicate}).

\section*{Data and code availability}

The code, fail-soft driver, scripts, and committed result files behind every reported value are public at \url{https://github.com/sbryngelson/ANEForge} under a permissive open-source license, archived as release \texttt{artifact-v1} at \url{https://doi.org/10.5281/zenodo.24210581}~\cite{aneforge_repo}.

\begin{acks}
During preparation of this manuscript the author used a large language model (Anthropic Claude) for prose and grammar polishing. The author reviewed and edited all such text and takes responsibility for the content.
\end{acks}

\clearpage
\bibliographystyle{ACM-Reference-Format}
\bibliography{references}

\clearpage
\appendix
\raggedbottom
\setcounter{table}{0}
\renewcommand{\thetable}{S\arabic{table}}
\section*{Supplementary material}
\addcontentsline{toc}{section}{Supplementary material}

\noindent These appendices are the paper's supplementary material.
Tables use an S prefix.
Unprefixed section and table references point to the main paper.

\section{Supplementary measurement tables}
\label{sec:appendix-tables}

This appendix reports the measurements summarized in the main text.

\begin{table}[htbp]
  \centering
  \small
  \caption{Comparison of idle-subtracted total-package active power from \texttt{powermetrics} and IOReport at five operating points over ten runs.
  Entries report medians and ten-run ranges.
  Both totals are ANE+GPU+CPU with the DRAM rail excluded (quantified separately below).
  Agreement measures cross-estimator consistency, not wall-meter accuracy.}
  \label{tab:ioreport}
  \begin{adjustbox}{max width=\linewidth}
\begin{tabular}{lccc}
    \toprule
    Operating point & powermetrics (W) & IOReport (W) & Agreement (median; 10-run range) \\
    \midrule
    GPU GEMM ($M=K=N=4096$, fp16) & 52.9 & 53.1 & \qty{99.3}{\percent} (93.0--99.9) \\
    CPU GEMM ($2048^2$, fp32, \qty{100}{\percent} CPU rail) & 16.8 & 16.5 & \qty{97.6}{\percent} (71.9--99.8) \\
    ANE stream (standalone relu, CPU-dispatch-heavy) & 8.3 & 8.3 & \qty{98.6}{\percent} (87.9--99.9) \\
    ANE attention (ViT block, $S=197$, $D=768$) & 7.9 & 8.2 & \qty{97.6}{\percent} (74.7--99.3) \\
    ANE conv (4-layer $3\times3$, $16\to64$ch @$32^2$) & 5.5 & 5.7 & \qty{96.9}{\percent} (73.2--99.5) \\
    \bottomrule
  \end{tabular}
\end{adjustbox}
\end{table}

IOReport lands within about \qty{2}{\percent} of \texttt{powermetrics} at the median on every point, in no systematic direction.
Two of the five points are CPU-dispatch-dominated (CPU GEMM, ANE stream), so the cross-check covers the attribution the total-package figure depends on.
The lowest-power ANE points carry the widest run-to-run spread, dipping to about \qty{73}{\percent} on occasional runs while their medians hold near \qty{97}{\percent}.

\begin{table}[htbp]
  \centering
  \small
  \caption{Silicon-versus-total-package perf/W (representative single-stream rows; idle-subtracted).
  Total-package perf/W includes CPU dispatch power.
  ANE-rail-only perf/W excludes it.
  Their difference is the measured host-dispatch share.}
  \label{tab:dual-bound}
  \begin{adjustbox}{max width=\linewidth}
\begin{tabular}{lrrr}
    \toprule
    Workload & total-package perf/W & ANE-rail-only perf/W & CPU-dispatch share \\
    \midrule
    conv ResNet-like ($C=256$, fused) & 2289 & 2677 & \qty{15}{\percent} \\
    GEMM compute ($K=4096$) & 751 & 1181 & \qty{37}{\percent} \\
    attention ViT ($S=197$) & 522 & 1074 & \qty{52}{\percent} \\
    GEMM bandwidth ($K=1024$) & 274 & 1007 & \qty{73}{\percent} \\
    conv single (tiny, dispatch-bound) & 25 & 176 & \qty{86}{\percent} \\
    \bottomrule
  \end{tabular}
\end{adjustbox}
\end{table}

\begin{table}[htbp]
  \centering
  \caption{Weight-stream effective bandwidth and energy, GEMV ($M=1$), $K=N$, fp16 (CPU $^*$fp32).
  GB/s is effective weight and I/O bytes divided by minimum latency.
  GB/s/W is GB/s divided by idle-subtracted active package power.
  The ANE reaches a \qtyrange{130}{143}{\giga\byte\per\second} plateau, and the GPU's unified-memory bandwidth is higher from $K=4096$ on, but the ANE moves bytes at higher energy efficiency at every $K$.}
  \label{tab:gemv-bw}
  \begin{adjustbox}{max width=\linewidth}
\begin{tabular}{rrrrrr}
    \toprule
    $K=N$ & ANE GB/s & GPU GB/s & CPU GB/s$^*$ & ANE GB/s/W & GPU GB/s/W \\
    \midrule
    512 & 6.1 & 3.7 & 277.6 & 0.8 & 0.4 \\
    1024 & 28.9 & 11.0 & 401.8 & 9.2 & 1.1 \\
    2048 & 75.6 & 32.4 & 63.6 & 13.0 & 4.9 \\
    4096 & 113.5 & 129.3 & 52.9 & \textbf{35.9} & 18.0 \\
    6144 & 129.8 & 169.1 & 62.7 & \textbf{53.1} & 21.1 \\
    8192 & 142.6 & 221.4 & 61.0 & \textbf{60.8} & 41.9 \\
    \bottomrule
  \end{tabular}
\end{adjustbox}
\end{table}

\begin{table}[htbp]
  \centering
  \caption{Compressed weight-stream speedup (single matmul, ANE via \texttt{aneforge}; median of five 200-eval blocks, batch 1 unless noted).
  Speedup is fp16 median latency divided by the compressed median.
  The bytes column is the compressed weight-blob fraction of fp16.
  int4 is per-tensor LUT.
  Sparse is the unstructured weight at about \qty{68}{\percent} zeros (about $0.38\times$ bytes).}
  \label{tab:compress-speedup}
  \begin{adjustbox}{max width=\linewidth}
\begin{tabular}{lrrr}
    \toprule
    Matmul ($K\times N$), batch & fp16 (ms) & int4 (bytes) & sparse (bytes) \\
    \midrule
    $1024\times1024$, $B=1$ & 0.095 & $0.94\times$ ($0.25\times$) & $0.94\times$ ($0.38\times$) \\
    $2048\times2048$, $B=1$ & 0.142 & $1.32\times$ & $1.38\times$ \\
    $4096\times4096$, $B=1$ & 0.312 & $1.65\times$ & $1.63\times$ \\
    $8192\times4096$, $B=1$ & 0.522 & $1.67\times$ & $1.82\times$ \\
    $256\times256$, $B=1$ & 0.089 & $0.71\times$ & $0.77\times$ \\
    $4096\times4096$, $B=64$ & 0.356 & $1.37\times$ & $1.52\times$ \\
    \bottomrule
  \end{tabular}
\end{adjustbox}
\end{table}

\begin{table}[htbp]
  \centering
  \footnotesize
  \setlength{\tabcolsep}{3pt}
  \caption{Fused-GPU baseline (M5 Pro, 6 s window, idle package = 1017 mW).
  GPU default versus \texttt{mx.compile}-fused versus ANE: latency (ms), idle-subtracted active package W, and perf/W (Gelem/s/W for the elementwise rows, GFLOP/s/W for attention, conv, and the stencil).
  Best latency and best perf/W per workload in \textbf{bold}.
  The stencil was measured in a separate session.
  Its active W is idle-subtracted and therefore comparable.
  Shapes: layer\_norm and gelu at $197\times768$; attention at $S=197$, $D=768$, $H=12$; conv stack of 16 layers, $64\to256$ channels, $32^2$ spatial, $k=3$; 5-point stencil on $256^2$ over 32 steps.}
  \label{tab:fused-gpu}
  \begin{adjustbox}{max width=\linewidth}
\begin{tabular}{lrrrrrrrrr}
    \toprule
    Workload & GPU def ms & fused ms & ANE ms & def W & fused W & ANE W & def perf/W & fused perf/W & ANE perf/W \\
    \midrule
    layer\_norm & 0.298 & \textbf{0.145} & 0.195 & 9.11 & 8.78 & \textbf{5.20} & 0.091 & 0.099 & \textbf{0.135} \\
    gelu & 0.222 & 0.120 & \textbf{0.088} & 8.53 & 7.32 & \textbf{7.37} & 0.100 & 0.129 & \textbf{0.204} \\
    attention & 0.270 & 0.288 & \textbf{0.233} & 15.07 & 15.10 & \textbf{7.64} & 212 & 212 & \textbf{566} \\
    conv stack & 3.166 & 3.164 & \textbf{0.763} & 31.89 & 31.93 & \textbf{9.16} & 180 & 180 & \textbf{2477} \\
    5-pt stencil & 1.323 & 1.281 & \textbf{0.132} & 33.02 & 33.16 & \textbf{7.28} & 0.86 & 0.86 & \textbf{38.3} \\
    \bottomrule
  \end{tabular}
\end{adjustbox}
\end{table}

\begin{table}[htbp]
  \centering
  \small
  \caption{Like-for-like fp16 GPU baseline for the three real models (torch-MPS \texttt{.half()}), paired per run against the ANE.
  Energy columns are means over $N{=}16$ independent cold starts.
  Each run re-samples its own idle baseline, shuffles the workload order, and is discarded if that baseline exceeds \qty{2}{\watt}.
  The advantage is a geometric mean of the per-run ratios with a \qty{95}{\percent} interval, aggregated in log space because a ratio is bounded below and right-skewed.
  The fp32 column is a single-session result without an interval; every other energy figure in this table is replicated.
  The compiled-MLX comparison for all three models is in the formulation-fairness appendix (\Cref{sec:formulation}).}
  \label{tab:real-fp16}
  \begin{adjustbox}{max width=\linewidth}
\begin{tabular}{lrrrrr}
    \toprule
    Model & GPU fp32 mJ & GPU fp16 mJ & GPU fp16 lat (ms) & ANE mJ (committed) & ANE energy adv.\ (fp16) \\
    \midrule
    ResNet-18 & 39.1 & 21.3 & 2.033 & 2.30 & $8.9\times$ $[7.5, 10.7]$ \\
    MiniLM    & 28.8 & 17.6 & 1.920 & 2.74 & $6.7\times$ $[5.9, 7.6]$ \\
    ViT-B/16  & 1382.7 & 577.9 & \textbf{15.92} & 65.7 & $9.4\times$ $[7.9, 11.1]$ \\
    \bottomrule
  \end{tabular}
\end{adjustbox}
\end{table}

Run-to-run variation is asymmetric.
The ANE arm's coefficient of variation is \qtyrange{11}{52}{\percent} against \qtyrange{1.7}{30}{\percent} for torch-MPS.
ResNet-18's torch-MPS energy alone spans \qtyrange{20}{49}{\milli\joule} across cold starts, so one GPU measurement does not determine the ratio.
Moving the GPU from fp32 to fp16 cuts its energy by \qtyrange{39}{58}{\percent}.

\begin{table}[htbp]
  \centering
  \caption{Vision, 6$\times$ $3\times3$ conv (16$\to$64ch, $32^2$), then a global average pool and a 100-way fully connected layer; items = images.}
  \label{tab:serving-vision}
  \begin{adjustbox}{max width=\linewidth}
\begin{tabular}{lrrrrrr}
    \toprule
    B & ANE thr & GPU thr & ANE perf/W & GPU perf/W & ANE mJ/it & GPU mJ/it \\
    \midrule
    1 & \textbf{11759} & 2067 & \textbf{1536} & 232 & \textbf{0.65} & 4.32 \\
    4 & \textbf{28768} & 9183 & \textbf{2904} & 367 & \textbf{0.34} & 2.72 \\
    16 & \textbf{50962} & 13984 & \textbf{4484} & 457 & \textbf{0.22} & 2.19 \\
    64 & \textbf{62204} & 16312 & \textbf{4607} & 466 & \textbf{0.22} & 2.14 \\
    256 & \textbf{66969} & 17338 & \textbf{4840} & 475 & \textbf{0.21} & 2.10 \\
    \bottomrule
  \end{tabular}
\end{adjustbox}
\end{table}

\begin{table}[htbp]
  \centering
  \caption{Encoder, pre-LN transformer block, $S=128$, $D=768$, $H=12$, $D_\text{ff}=3072$; items = sequences.}
  \label{tab:serving-encoder}
  \begin{adjustbox}{max width=\linewidth}
\begin{tabular}{lrrrrrr}
    \toprule
    B & ANE thr & GPU thr & ANE perf/W & GPU perf/W & ANE mJ/it & GPU mJ/it \\
    \midrule
    1 & \textbf{2756} & 1262 & \textbf{374} & 69 & \textbf{2.67} & 14.51 \\
    4 & \textbf{2984} & 2546 & \textbf{391} & 103 & \textbf{2.56} & 9.73 \\
    16 & \textbf{2786} & 2584 & \textbf{402} & 129 & \textbf{2.48} & 7.73 \\
    64 & 2364 & \textbf{2947} & \textbf{414} & 145 & \textbf{2.41} & 6.88 \\
    256 & 1031 & \textbf{2948} & \textbf{285} & 150 & \textbf{3.51} & 6.65 \\
    \bottomrule
  \end{tabular}
\end{adjustbox}
\end{table}

\begin{table}[htbp]
  \centering
  \caption{Attention, self-attention block, $S=128$, $D=768$, $H=12$; items = sequences; decomposed scaled-dot-product attention (SDPA), true rank-4 batch.
  $^\dagger$The $B=4$ GPU point did not reproduce.
  An independent run measured about \num{11000} it/s.
  The replicated throughput crossover is $B^*=2.23$.}
  \label{tab:serving-attention}
  \begin{adjustbox}{max width=\linewidth}
\begin{tabular}{lrrrrrr}
    \toprule
    B & ANE thr & GPU thr & ANE perf/W & GPU perf/W & ANE mJ/it & GPU mJ/it \\
    \midrule
    1 & \textbf{6842} & 1948 & \textbf{715} & 297 & \textbf{1.40} & 3.37 \\
    4 & 8972 & 4238$^\dagger$ & \textbf{1114} & 495 & \textbf{0.90} & 2.02 \\
    16 & 9778 & \textbf{21888} & \textbf{1122} & 580 & \textbf{0.89} & 1.72 \\
    64 & 7782 & \textbf{25463} & \textbf{1044} & 631 & \textbf{0.96} & 1.58 \\
    256 & 8088 & \textbf{24880} & \textbf{1027} & 618 & \textbf{0.97} & 1.62 \\
    \bottomrule
  \end{tabular}
\end{adjustbox}
\end{table}

\begin{table}[htbp]
  \centering
  \caption{Batched GEMM, \texttt{[B,128,1024] @ [1024,1024]}; items = batch elements.}
  \label{tab:serving-gemm}
  \begin{adjustbox}{max width=\linewidth}
\begin{tabular}{lrrrrrr}
    \toprule
    B & ANE thr & GPU thr & ANE perf/W & GPU perf/W & ANE mJ/it & GPU mJ/it \\
    \midrule
    1 & \textbf{8276} & 3734 & \textbf{736} & 382 & 1.36 & 2.62 \\
    4 & \textbf{18540} & 11228 & \textbf{1508} & 716 & 0.66 & 1.40 \\
    16 & 23495 & \textbf{54748} & \textbf{1903} & 1841 & 0.53 & 0.54 \\
    64 & 27198 & \textbf{94407} & 2135 & \textbf{2148} & 0.47 & 0.47 \\
    256 & 27577 & \textbf{109589} & \textbf{2223} & 2063 & 0.45 & 0.48 \\
    \bottomrule
  \end{tabular}
\end{adjustbox}
\end{table}

The three device maps that follow share the columns and conventions of the main paper's device map.
Best speed and best perf/W per row are in \textbf{bold}.
Each caption states only the conditions and omissions specific to its host.

\begin{table}[htbp]
  \centering
  \footnotesize
  \setlength{\tabcolsep}{2pt}
  \caption{Single-stream device map on the M1 Max host (h13 ANE; 6 s power window, idle package = 2014 mW).
  The three real-model classes were skipped on this host (torchvision and transformers unavailable), so the table reports the eleven synthetic classes.}
  \label{tab:m1-map}
  \begin{adjustbox}{max width=\linewidth}
\begin{tabular}{llrrllll}
    \toprule
    Class & Workload & ANE lat & GPU lat & ANE W (CV) & GPU W (CV) & ANE perf/W & GPU perf/W \\
    \midrule
    GEMM floor & M64$\cdot$K256$\cdot$N256 & \textbf{0.164} & 0.184 & 2.50 (\qty{16}{\percent}) & 1.23 (\qty{8}{\percent}) & 17.0 GF/W & \textbf{27.1} GF/W \\
    GEMM bandwidth & M128$\cdot$K1024$\cdot$N1024 & 0.405 & \textbf{0.242} & 1.92 (\qty{5}{\percent}) & 3.51 (\qty{32}{\percent}) & \textbf{470} GF/W & 207 GF/W \\
    GEMM compute & M256$\cdot$K4096$\cdot$N4096 & 2.394 & \textbf{1.179} & 4.42 (\qty{11}{\percent}) & 32.47 (\qty{1}{\percent}) & \textbf{775} GF/W & 192 GF/W \\
    conv & single $3{\times}3$ ($16\to64$) & \textbf{0.191} & 0.211 & 3.94 (\qty{4}{\percent}) & 3.43 (\qty{11}{\percent}) & \textbf{23.2} GF/W & 17.7 GF/W \\
    conv & ResNet-like $3{\times}3{\times}16$ ($64\to256$) & \textbf{1.918} & 3.944 & 4.35 (\qty{3}{\percent}) & 30.09 (\qty{6}{\percent}) & \textbf{2063} GF/W & 142 GF/W \\
    attention & ViT block $S=197$ & 0.613 & \textbf{0.573} & 4.48 (\qty{8}{\percent}) & 4.04 (\qty{8}{\percent}) & \textbf{370} GF/W & 187 GF/W \\
    attention & long-seq $S=512$ & 1.815 & \textbf{0.855} & 3.54 (\qty{15}{\percent}) & 9.82 (\qty{23}{\percent}) & \textbf{544} GF/W & 217 GF/W \\
    norm & layer\_norm (197,768) & 0.321 & \textbf{0.308} & 4.39 (\qty{14}{\percent}) & 3.84 (\qty{13}{\percent}) & 9.09e7/W & \textbf{1.01e8}/W \\
    norm & rms\_norm (197,768) & 1.311 & \textbf{0.255} & 3.51 (\qty{10}{\percent}) & 4.06 (\qty{15}{\percent}) & 6.29e7/W & \textbf{1.07e8}/W \\
    norm & group\_norm (1,256,$64^2$) & 1.025 & \textbf{0.358} & 4.66 (\qty{9}{\percent}) & 2.54 (\qty{27}{\percent}) & 3.49e8/W & \textbf{6.14e8}/W \\
    sci & 5-pt stencil $256^2$, 32 steps & \textbf{0.223} & 2.066 & 4.46 (\qty{5}{\percent}) & 26.06 (\qty{2}{\percent}) & \textbf{33.8} GF/W & 0.60 GF/W \\
    \bottomrule
  \end{tabular}
\end{adjustbox}
\end{table}

\begin{table}[htbp]
  \centering
  \footnotesize
  \setlength{\tabcolsep}{2pt}
  \caption{Single-stream device map on the M2 Pro host (h14 ANE, the A14-generation engine; 6 s power window, idle package = 57 mW).
  The bottom block reports the three real models.
  There the GPU baseline is torch-MPS fp32 (marked $^*$, not a like-for-like fp16 comparison) and the last two columns report energy in mJ/inference instead of perf/W.
  MiniLM has no torch-MPS reference in the harness.}
  \label{tab:m2-map}
  \begin{adjustbox}{max width=\linewidth}
\begin{tabular}{llrrllll}
    \toprule
    Class & Workload & ANE lat & GPU lat & ANE W (CV) & GPU W (CV) & ANE perf/W & GPU perf/W \\
    \midrule
    GEMM floor & M64$\cdot$K256$\cdot$N256 & \textbf{0.110} & 0.206 & 2.66 (\qty{7}{\percent}) & 3.04 (\qty{5}{\percent}) & 11.4 GF/W & \textbf{16.6} GF/W \\
    GEMM bandwidth & M128$\cdot$K1024$\cdot$N1024 & \textbf{0.135} & 0.222 & 5.57 (\qty{2}{\percent}) & 7.78 (\qty{10}{\percent}) & \textbf{298} GF/W & 130 GF/W \\
    GEMM compute & M256$\cdot$K4096$\cdot$N4096 & 2.043 & \textbf{1.858} & 5.61 (\qty{7}{\percent}) & 21.25 (\qty{1}{\percent}) & \textbf{773} GF/W & 214 GF/W \\
    conv & single $3{\times}3$ ($16\to64$) & \textbf{0.107} & 0.219 & 2.93 (\qty{3}{\percent}) & 3.37 (\qty{7}{\percent}) & 25.1 GF/W & \textbf{31.4} GF/W \\
    conv & ResNet-like $3{\times}3{\times}16$ ($64\to256$) & \textbf{1.265} & 5.126 & 6.53 (\qty{1}{\percent}) & 20.52 (\qty{1}{\percent}) & \textbf{2235} GF/W & 173 GF/W \\
    attention & ViT block $S=197$ & \textbf{0.589} & 0.643 & 5.52 (\qty{11}{\percent}) & 12.88 (\qty{11}{\percent}) & \textbf{481} GF/W & 120 GF/W \\
    attention & long-seq $S=512$ & 1.253 & \textbf{1.144} & 4.63 (\qty{4}{\percent}) & 17.74 (\qty{8}{\percent}) & \textbf{567} GF/W & 150 GF/W \\
    norm & layer\_norm (197,768) & 0.283 & \textbf{0.278} & 4.30 (\qty{7}{\percent}) & 3.71 (\qty{8}{\percent}) & 1.08e8/W & \textbf{1.75e8}/W \\
    norm & rms\_norm (197,768) & 1.080 & \textbf{0.182} & 3.40 (\qty{10}{\percent}) & 3.47 (\qty{7}{\percent}) & 7.47e7/W & \textbf{2.19e8}/W \\
    norm & group\_norm (1,256,$64^2$) & 0.679 & \textbf{0.288} & 5.36 (\qty{3}{\percent}) & 5.72 (\qty{9}{\percent}) & 4.12e8/W & \textbf{5.80e8}/W \\
    sci & 5-pt stencil $256^2$, 32 steps & \textbf{0.371} & 3.349 & 4.21 (\qty{8}{\percent}) & 14.13 (\qty{2}{\percent}) & \textbf{37.6} GF/W & 0.79 GF/W \\
    \midrule
    model & ResNet-18 ($1\times3\times224^2$) & \textbf{1.064} & 3.379$^*$ & 4.19 (\qty{8}{\percent}) & 3.75 (\qty{26}{\percent}) & \textbf{2.6 mJ} & 22.3 mJ \\
    model & MiniLM encoder (1 sent) & \textbf{1.753} & -- & 3.73 (\qty{9}{\percent}) & -- & \textbf{3.0 mJ} & -- \\
    model & ViT-B/16 (12-layer, 197 tok) & \textbf{21.34} & 37.28$^*$ & 3.14 (\qty{1}{\percent}) & 18.45 (\qty{4}{\percent}) & \textbf{67.9 mJ} & 714.3 mJ \\
    \bottomrule
  \end{tabular}
\end{adjustbox}
\end{table}

\begin{table}[t]
  \centering
  \footnotesize
  \setlength{\tabcolsep}{2pt}
  \caption{Single-stream device map on the M4 host (h16 ANE, the base-M4 engine; 6 s power window, idle package = 135 mW).
  The three real-model rows are absent because ResNet-18 hung at compile on h16.
  The synthetic map provides the h16 comparison, and reproduces to within a few percent across two independent runs.}
  \label{tab:m4-map}
  \begin{adjustbox}{max width=\linewidth}
\begin{tabular}{llrrllll}
    \toprule
    Class & Workload & ANE lat & GPU lat & ANE W (CV) & GPU W (CV) & ANE perf/W & GPU perf/W \\
    \midrule
    GEMM floor & M64$\cdot$K256$\cdot$N256 & \textbf{0.076} & 0.202 & 1.31 (\qty{9}{\percent}) & 2.46 (\qty{7}{\percent}) & \textbf{27.9} GF/W & 18.1 GF/W \\
    GEMM bandwidth & M128$\cdot$K1024$\cdot$N1024 & \textbf{0.184} & 0.284 & 4.09 (\qty{6}{\percent}) & 6.31 (\qty{7}{\percent}) & \textbf{407} GF/W & 133 GF/W \\
    GEMM compute & M256$\cdot$K4096$\cdot$N4096 & \textbf{2.117} & 2.692 & 3.94 (\qty{5}{\percent}) & 14.90 (\qty{12}{\percent}) & \textbf{1020} GF/W & 209 GF/W \\
    conv & single $3{\times}3$ ($16\to64$) & \textbf{0.092} & 0.198 & 3.51 (\qty{5}{\percent}) & 2.63 (\qty{5}{\percent}) & \textbf{57.9} GF/W & 32.1 GF/W \\
    conv & ResNet-like $3{\times}3{\times}16$ ($64\to256$) & \textbf{1.158} & 6.045 & 5.65 (\qty{5}{\percent}) & 15.37 (\qty{6}{\percent}) & \textbf{2746} GF/W & 197 GF/W \\
    attention & ViT block $S=197$ & \textbf{0.334} & 0.762 & 4.26 (\qty{5}{\percent}) & 9.89 (\qty{6}{\percent}) & \textbf{780} GF/W & 122 GF/W \\
    attention & long-seq $S=512$ & \textbf{1.100} & 1.478 & 3.36 (\qty{3}{\percent}) & 11.10 (\qty{9}{\percent}) & \textbf{868} GF/W & 186 GF/W \\
    norm & layer\_norm (197,768) & 0.285 & \textbf{0.275} & 2.62 (\qty{5}{\percent}) & 3.56 (\qty{6}{\percent}) & \textbf{2.04e8}/W & 1.66e8/W \\
    norm & rms\_norm (197,768) & 0.231 & \textbf{0.221} & 4.05 (\qty{4}{\percent}) & 3.06 (\qty{6}{\percent}) & \textbf{2.78e8}/W & 2.10e8/W \\
    norm & group\_norm (1,256,$64^2$) & 0.661 & \textbf{0.336} & 4.31 (\qty{5}{\percent}) & 5.45 (\qty{6}{\percent}) & 4.02e8/W & \textbf{5.07e8}/W \\
    sci & 5-pt stencil $256^2$, 32 steps & \textbf{0.139} & 3.638 & 3.00 (\qty{5}{\percent}) & 9.77 (\qty{6}{\percent}) & \textbf{77.6} GF/W & 1.02 GF/W \\
    \bottomrule
  \end{tabular}
\end{adjustbox}
\end{table}

\paragraph{Stride aliasing on h16.}
A dense weight-stream $K$ sweep on h16 shows a dip to about \qty{25}{\giga\byte\per\second} at exactly the powers of two $K=4096$, $8192$, and $16384$.
Every neighboring size, including $K=12288$, streams at full rate.
Neither the M2 Pro nor the M5 Pro shows the dip, and every point still exceeds the activation ceiling by at least $2.4\times$.

\paragraph{Per-class cross-generation agreement.}
Counting against the M5 reference map (the main paper's device map) across the eleven synthetic classes in Tables~S11--S13, the faster device is unchanged on 7 and the more efficient device on 7 (3 classes on both axes).
The perf/watt disagreements all fall in sub-millisecond dispatch-bound and tiny-norm rows, that is, outside the substantial classes above the dispatch floor that the main paper defines.
They are GEMM-floor perf/watt (GPU on both h13 and h14), single-conv perf/watt (GPU on h14), and layer\_norm and rms\_norm perf/watt (GPU on both).
On those rows package power is dominated by host dispatch.
The speed disagreements are the GEMM and attention rows, which switch to the GPU only on the older parts and back to the ANE on the newest (h16).

\section{FLOP and byte accounting per archetype}
\label{sec:flop-formulas}

Every throughput, perf/watt, and arithmetic-intensity figure in this paper derives from the FLOP and byte formulas below, which are taken directly from the measurement scripts.
Three conventions apply throughout.
(i)~Multiply--accumulates are counted as 2~FLOPs.
(ii)~Element traffic is priced at 2~B (fp16) for the GPU and ANE and 4~B (fp32) for the CPU, so every CPU AI is half the fp16 value listed.
(iii)~Elementwise transcendentals are charged a flat per-element allowance: 1~FLOP/element for relu/copy-class ops, 5~FLOPs/element for softmax (max, subtract, exp, sum, divide), and 10~FLOPs/element for gelu/silu (the erf/tanh polynomial approximation) and for layer\_norm (mean, variance, normalize, scale, shift).
These are order-of-magnitude conventions, not instruction counts.
Scaling any allowance changes the derived FLOP rate but not the measured bandwidth or the regime classification.
The memory-bound archetypes are measured in bytes per second and reported in FLOP-denominated units only through these factors, and no tested archetype crosses a ridge.

\begin{table}[htbp]
  \centering
  \footnotesize
  \setlength{\tabcolsep}{2pt}
  \caption{Roofline-placement archetypes (the main paper's roofline-placement table): FLOP
  formula, fp16 byte formula, and resulting arithmetic intensity.
  $b$ denotes bytes per element (2 for fp16; the CPU's $b=4$ halves each AI).
  For the memory-bound rows the achieved GFLOP/s in the main paper's roofline-placement table is derived as AI~$\times$~measured GB/s; for the compute-bound rows it is the size-specific measurement.
  Attention counts $QK^\top$ and $AV$ at $2HS^2d_h$ FLOPs each with softmax at 5 FLOPs per score; its bytes are $Q$, $K$, $V$, and the output ($4HSd_h$) plus the score matrix written and re-read for softmax ($2HS^2$, the standalone/spilled convention).}
  \label{tab:flop-formulas-roofline}
  \begin{adjustbox}{max width=\linewidth}
\begin{tabular}{lllr}
    \toprule
    Archetype & FLOPs & Bytes (fp16) & AI (FLOP/B) \\
    \midrule
    matmul square $N=512$   & $2N^3$ & $3N^2 b$ \enspace (AI $= N/3$) & 170.7 \\
    matmul square $N=2048$  & $2N^3$ & $3N^2 b$ & 682.7 \\
    matmul square $N=8192$  & $2N^3$ & $3N^2 b$ & 2730.7 \\
    GEMV / decode ($M=1$, $K=N=4096$) & $2MKN = 33.6\,$MF & $(MK + KN + MN)\,b \approx KN\,b$ & 1.0 \\
    conv $3{\times}3$ ($C=64$, $B=16$, $64^2$) & $2BC_oC_i\,9HW = 4.83\,$GF & $(BC_iHW + C_oC_i 9 + BC_oHW)\,b$ & 286.7 \\
    conv $3{\times}3$ ($C=512$, $B=4$, $64^2$) & $2BC_oC_i\,9HW = 77.3\,$GF & $(BC_iHW + C_oC_i 9 + BC_oHW)\,b$ & 2019.9 \\
    attention ($S=512$, $d_h=64$, $H=12$) & $4HS^2d_h + 5HS^2 = 0.821\,$GF & $(4HSd_h + 2HS^2)\,b$ & 52.2 \\
    softmax & $5$ /elem & $2$ elem/elem $\times\, b$ & 1.25 \\
    gelu / silu & $10$ /elem & $2$ elem/elem $\times\, b$ & 2.5 \\
    layer\_norm & $10$ /elem & $2$ elem/elem $\times\, b$ & 2.5 \\
    relu / copy & $1$ /elem & $2$ elem/elem $\times\, b$ & 0.25 \\
    \midrule
    matmul int4 ($4096^2$, $B=1$)   & $2BKN = 33.6\,$MF & $KN/2 + (BK{+}BN)\,b$ & 4.0 \\
    matmul int4 ($4096^2$, $B=128$) & $2BKN = 4.29\,$GF & $KN/2 + (BK{+}BN)\,b$ & 409.6 \\
    fused 3-conv ($C=256$, $B=4$, $64^2$) & $3 \cdot 2BC^2 9HW = 58.0\,$GF & fused in/out $+\, 3 \cdot 9C^2$; standalone $3\times$ & 2854 / 1076 \\
    \bottomrule
  \end{tabular}
\end{adjustbox}
\end{table}

\begin{table}[htbp]
  \centering
  \footnotesize
  \setlength{\tabcolsep}{3pt}
  \caption{Device-map workload classes (the main paper's device map): the exact
  work count used to convert the measured latency and active power into throughput and perf/watt.
  The norm family is counted in elements normalized (Gelem/s), not FLOPs; the real models report energy per inference (mJ/inf) with no FLOP count.
  AI is shown for the fp16 fused-program traffic model of \Cref{tab:flop-formulas-roofline} (input $+$ weights $+$ output; on-chip intermediates uncounted) and is derived here, not computed by the script.
  The ViT attention count is the four $D{\times}D$ projections ($8SD^2$) plus $QK^\top$ and $AV$ ($4HS^2d_h$, $d_h = D/H$), with in-graph softmax FLOPs uncounted (conservative); the stencil is counted as the dense $3{\times}3$ convolution executed, per the main paper's device map caption.}
  \label{tab:flop-formulas-devicemap}
  \begin{adjustbox}{max width=\linewidth}
\begin{tabular}{lllr}
    \toprule
    Device-map class & Work formula (per call) & Value & AI (fp16) \\
    \midrule
    GEMM floor ($M=64,K=N=256$)       & $2MKN$ & $8.39$\,MF & 42.7 \\
    GEMM bandwidth ($M=128,K=N=1024$) & $2MKN$ & $268.4$\,MF & 102.4 \\
    GEMM compute ($M=256,K=N=4096$)   & $2MKN$ & $8.59$\,GF & 227.6 \\
    conv single $3{\times}3$ ($16\to64$, $32^2$) & $2\,C_iC_o 9HW$ & $18.9$\,MF & 103.5 \\
    conv ResNet-like $\times16$ ($64\to256$, then $256^2{\times}15$, $32^2$) & $\sum_{d=0}^{15} 2\,C_i^{(d)}C_o 9HW$ & $18.42$\,GF & 988 \\
    attention ViT ($S=197,D=768,H=12$) & $8SD^2 + 4HS^2d_h$ & $1.049$\,GF & -- \\
    attention long-seq ($S=512$) & $8SD^2 + 4HS^2d_h$ & $3.221$\,GF & -- \\
    layer\_norm / rms\_norm $(197,768)$ & elements $= \prod(\text{shape})$ & $0.151$\,Melem & 2.5 \\
    group\_norm $(1,256,64^2)$ & elements & $1.049$\,Melem & 2.5 \\
    5-pt stencil ($256^2$, 32 steps) & $\text{steps} \times 2 \cdot 9 \cdot HW$ & $37.7$\,MF & -- \\
    ResNet-18 / MiniLM / ViT-B/16 & energy per inference & mJ/inf & -- \\
    \bottomrule
  \end{tabular}
\end{adjustbox}
\end{table}

\begin{table}[htbp]
  \centering
  \footnotesize
  \caption{Bytes-moved models for the bandwidth roofline: achieved bandwidth $=$ bytes/min-latency with the stated traffic factor; $N$ is the element count and $dt$ the dtype size (2~B fp16 for GPU/ANE, 4~B fp32 for the CPU, whose proxy op is a memory-bound scalar multiply or copy, never a transcendental).
  The saturation-sweep FLOP counts are $2N^3$ for square GEMM and $2BC^2\,9HW$ for $3{\times}3$ conv at $64{\times}64$ spatial.}
  \label{tab:byte-models}
  \begin{adjustbox}{max width=\linewidth}
\begin{tabular}{llp{32em}}
    \toprule
    Bandwidth type & Bytes moved & Reason \\
    \midrule
    streaming (relu / $x{\cdot}2$) & $2N\,dt$ & read $N$ $+$ write $N$; the primary bandwidth-roof probe \\
    gelu (light compute) & $2N\,dt$ & same traffic; arithmetic does not add bytes \\
    reduction (sum, last axis) & $1N\,dt$ & read $N$, write $\sim N/D \approx 0$ \\
    layer\_norm (last axis) & $2N\,dt$ & read-once/write-once, statistics on-chip \\
    softmax (last axis) & $2N\,dt$ & logical two passes (max+exp, sum+div) are fused; read-once/write-once is paid DRAM traffic \\
    \bottomrule
  \end{tabular}
\end{adjustbox}
\end{table}

\section{A silicon-versus-package perf/watt bound}
\label{sec:silicon-bound}

The headline perf/watt divides throughput by total-package active power, which charges the ANE for its own CPU dispatch.
An ANE program still draws several watts of host CPU power submitting work.
The per-rail breakdown puts that CPU share between \qty{15}{\percent} on a compute-heavy fused conv and \qty{86}{\percent} on a single tiny op.
We therefore report two figures from the same per-rail data: total-package perf/watt, which charges the ANE the full dispatch power, and ANE-rail-only perf/watt, which excludes it.

On the device-map classes of the main paper the ANE is more efficient on the total-package figure (\Cref{tab:dual-bound}), and the rail-only figure is $1.2$ to $4\times$ larger.
The regime where the CPU dominates the package draw is the single tiny op at \qtyrange{80}{86}{\percent} CPU, where standalone dispatch leaves the engine mostly idle and the rail-only column reads 176 against 25 GFLOP/s/W.
We keep total-package as the headline because it is the marginal cost a deployer pays today, and report the rail-only figure alongside it.

\section{Silicon versus toolchain: a fused-GPU baseline}
\label{sec:fused-gpu}

The single-stream device map compares \texttt{aneforge}'s fused ANE programs against MLX's default (un-fused) GPU dispatch.
Part of the measured gap could therefore be MLX's kernel-launch overhead and not silicon.
Fusing the GPU with \texttt{mx.compile} tests this.
Five representative workloads (a ViT-sized layer\_norm and gelu, a self-attention block, a 16-deep conv stack, and the 32-step stencil that carries the device map's largest perf/watt gap) run on the GPU in both default and \texttt{mx.compile}-fused form, against the ANE, under the same per-engine energy harness.
\Cref{tab:fused-gpu} reports them.

Fusion moves some rows and not others.
It helps the multi-pass elementwise ops: layer\_norm speeds up $2.05\times$ and gelu $1.85\times$ with \texttt{mx.compile}, and on one workload it reverses the within-session ordering: fused-MLX layer\_norm latency (\qty{0.145}{\milli\second}) is below the ANE's (\qty{0.195}{\milli\second}), where this session's default-MLX baseline (\qty{0.298}{\milli\second}) had been slower than the ANE.
The device-map session measured default MLX at \qty{0.176}{\milli\second}, so the small elementwise-latency rows are both session- and toolchain-sensitive.

The ANE has the higher perf/watt on all five workloads, because it draws far less power (conv at about \qty{9}{\watt} versus about \qty{32}{\watt}, attention at about \qty{7.6}{\watt} versus about \qty{15}{\watt}), so even where the GPU matches or improves on its latency, the ANE does the work at a fraction of the energy.
Fusion does not help the compute-bound or the bandwidth-bound multi-pass workloads: attention is $0.94\times$ (fusion slightly slows it), the conv stack is $1.00\times$, and the stencil is only $1.03\times$, because those are either already single fused MLX kernels or, in the stencil's case, a chain whose per-step cost is the bandwidth of streaming a $256\times256$ field, not kernel-launch overhead that fusion could remove.
The large gaps are unchanged under fusion: the ANE is $4.1\times$ faster than even fused-MLX on the conv stack, $1.2\times$ faster on attention, and $9.7\times$ faster on the stencil.
Fusing the GPU barely moves the stencil ($1.03\times$), so kernel-launch overhead does not explain that row.
\Cref{sec:formulation} shows the formulation does: the dense $3\times3$ convolution is a poor GPU expression of a 5-point stencil, and replacing it moves that row from $43.8\times$ to $4.7\times$.
These results exclude kernel-launch overhead as the main cause of the tested gaps.
They do not establish formulation fairness, and a fused baseline is not by itself an optimized one.

\section{Replicate and variance protocol}

The within-window coefficient of variation (CV\%) reported per measurement in the main paper does not measure variation across independent executions.
We therefore repeated the contested verdicts in fresh processes: the conv and GEMM perf/watt comparisons at peak, the large-margin ANE advantages, and the two serving crossovers.
Every run re-samples its own idle baseline and randomizes the workload and device order (ANE-first versus GPU-first) to break warm-up and ordering bias.
The two compute-bound rows were run separately, one workload per process under an inter-run cooldown, so no prior workload pre-heats the die.
Both devices are measured within the same run and share its idle baseline and ambient state, so the run is the pairing unit and inference is on the per-run differences.

We test each difference with a two-sided paired $t$-test and report the mean difference with a $t$-based \qty{95}{\percent} confidence interval (CI).
Equivalence is not inferred from failure to reject a difference.
Near-equivalent rows are tested with two one-sided tests (TOST) against a declared margin of $\pm\qty{5}{\percent}$ of the pooled device mean, and equivalence is declared only when both reject at $\alpha=0.05$.
Simultaneous difference verdicts are corrected with Holm--Bonferroni, so a difference requires an adjusted $p<0.05$ and a CI excluding zero.
Rows establishing neither a difference nor $\pm\qty{5}{\percent}$ equivalence are reported as indeterminate, and the two serving-crossover rows are estimation problems that carry confidence intervals only.
At these sample sizes an equivalence verdict is bounded by its CI, and margins much below a few percent cannot be certified.

Run-to-run idle-baseline drift is the dominant variance source.
The per-run idle package draw varies by several hundred milliwatts, and a run whose idle baseline is corrupted by transient background activity is gated out.
A run is dropped when its idle package power exceeds $\max(\qty{2}{\watt},\,2\times$ median idle$)$.
The criterion is evaluated on the idle sample alone, before the workload is measured, so it is blind to the outcome it could affect.
It is protective only while most runs are clean: the median rises with widespread contamination, and the $\max$ then widens the threshold instead of tightening it.
The real-model replication of \Cref{tab:real-fp16} therefore uses an absolute \qty{2}{\watt} threshold.
Background activity during the idle sample raises the baseline, over-subtracts, and inflates perf/watt, so the corrupted run is treated as a measurement fault, not a sample of the quantity of interest.
One run of the ten on the saturation pair exceeded the gate.
On the compute-bound GEMM row its exclusion tightens the statistics (ANE SD from 456 to 88, paired-$t$ $p$ from $0.019$ to $8.9\times10^{-6}$, minimum detectable effect from \qty{56.5}{\percent} to \qty{11.9}{\percent} of pooled) and reduces the estimated ANE advantage from $387.6$ to $253.2$.
On \texttt{conv\_C512} it moves an indeterminate verdict to one favoring the GPU (mean difference $-70.1$, CI $[-230.1, 89.9]$ ungated, $-123.7$, CI $[-242.6, -4.8]$ gated).
The gate therefore improves precision without systematically favoring the ANE.
\Cref{tab:replicate} reports the gated statistics, and the artifact retains both variants.
The absolute idle floor differs between sessions, for example \qty{1196}{\milli\watt} in the device-map run against \qty{1017}{\milli\watt} in the fused-GPU run, so every run re-samples its own baseline instead of reusing a stored one.
Every perf/watt number is idle-subtracted, so this drift propagates into the across-run standard deviation (SD), which is the variance figure we quote.
Thermal pressure remained Nominal throughout, but this M5 Pro exposes only the OS thermal-pressure level through \texttt{powermetrics} and not a die temperature, so that rules out a pressure-level throttle and not a sub-threshold drift in clocks.

\begin{table}[htbp]
  \centering
  \footnotesize
  \caption{Per-claim replication of the contested verdicts of the main paper's device-map, ceiling, and serving sections.
  Each claim is measured over independent cold-start runs (idle re-sampled and outlier-gated per run, workload and device order randomized, thermal pressure Nominal throughout); the two compute-bound rows are measured in isolation under inter-run cooldown so no workload pre-heats the die.
  perf/W is idle-subtracted; values are mean $\pm$ across-run SD; Diff is the paired ANE--GPU mean with a paired-$t$ \qty{95}{\percent} CI; $p$ is two-sided paired $t$, Holm-adjusted across the difference tests.
  A difference requires the CI to exclude 0 at adjusted $p<0.05$; equivalent requires two one-sided tests within the $\pm\qty{5}{\percent}$ margin at $\alpha=0.05$; crossover rows are estimation-only.}
  \label{tab:replicate}
  \setlength{\tabcolsep}{2pt}
  \begin{adjustbox}{max width=\linewidth}
\begin{tabular}{llllll}
    \toprule
    Claim (perf/W) & ANE mean $\pm$ SD & GPU mean $\pm$ SD & Diff [\qty{95}{\percent} CI] & $p$ (Holm) & Verdict \\
    \midrule
    conv @ $C=512$ (GF/W) & $1352 \pm 49$ & $1427 \pm 38$ & $-74$ $[-94, -55]$ & $<0.001$ & \textbf{GPU diff.\ ($\sim$\qty{5}{\percent})} \\
    GEMM @ serving $B=64$ (items/s/W) & $2075 \pm 34$ & $1945 \pm 5$ & $+130$ $[+87, +174]$ & $0.003$ & \textbf{ANE difference} \\
    conv ResNet-like (GF/W) & $2253 \pm 77$ & $172 \pm 2$ & $+2081$ $[+1988, +2173]$ & $<0.001$ & \textbf{ANE difference} \\
    attention $S=197$ (GF/W) & $517 \pm 12$ & $190 \pm 4$ & $+326$ $[+316, +337]$ & $<0.001$ & \textbf{ANE difference} \\
    GEMM-compute $K=4096$ (GF/W) & $811 \pm 88$ & $558 \pm 15$ & $+253$ $[+194, +312]$ & $<0.001$ & \textbf{ANE difference} \\
    Attention crossover $B^*$ & $2.23 \pm 0.02$ & -- & CI $[2.20, 2.25]$ & -- & \textbf{Stable} (est.\ only) \\
    Encoder crossover $B^*$ & $31.1 \pm 1.3$ & -- & CI $[29.5, 32.7]$ & -- & \textbf{Stable} (est.\ only) \\
    \bottomrule
  \end{tabular}
\end{adjustbox}
\end{table}

\section{CPU (fp32/AMX) reference figures}
\label{sec:cpu-leg}

The CPU enters the main paper as the fp32 accuracy reference and as a contextual roofline comparand, not as a like-for-like fp16 device.
This appendix collects the CPU figures that the body tables omit so the comparison stays focused on the ANE versus the GPU.
The CPU runs numpy/Accelerate (fp32, via Apple's AMX matrix coprocessor): it moves twice the bytes per element of the fp16 devices, halving its arithmetic intensities, so it is the least-comparable of the rooflines and is fastest only on tiny dispatch-bound shapes where call overhead, not FLOPs, dominates.
Its conv roof has two values, the \num{11} GFLOP/s im2col path the saturation harness defaults to and a tuned native conv (torch, fp32, the same shape) at \num{2077} GFLOP/s; the im2col figure is an implementation floor, and power was not separately metered for the tuned path.
The CPU GEMM power is flat at about \qty{16}{\watt} from $N=512$ to $N=8192$ while throughput holds at about \qty{1.8}{\tera\flop\per\second}, the signature of an AMX-bound ceiling~\cite{amx_re}.
On the roofline placement the CPU GEMV (\qty{124}{\percent} of roof) and gelu (\qty{135}{\percent}) entries are cache residency: at those working sets the CPU streams from cache at \qtyrange{277}{402}{\giga\byte\per\second}, far above its DRAM-derived roof, the phenomenon the cache-aware roofline~\cite{ilic2014carm} models.

\begin{table}[htbp]
  \centering
  \small
  \caption{CPU (fp32, Accelerate/AMX) reference figures removed from the main paper's body tables, M5 Pro.
  Compute peaks (conv im2col), bandwidth ceilings, and end-to-end decode throughput, as a contextual fp32 reference.
  The CPU is not directly comparable to the fp16 GPU/ANE columns (its AI is the fp32 value, half the fp16 AI).}
  \label{tab:cpu-leg}
  \begin{adjustbox}{max width=\linewidth}
\begin{tabular}{lll}
    \toprule
    Quantity & CPU value & Note \\
    \midrule
    GEMM peak GFLOP/s            & 1\,899 ($N=2048$), relerr 8.1e-7 & 112 GFLOP/s/W ($N=2048$) \\
    conv peak GFLOP/s            & 11 (im2col) & 1.05 GFLOP/s/W \\
    \midrule
    streaming GB/s              & 130.4 & 9.04 GB/s/W \\
    reduction sum GB/s          & 41.7  & 2.95 GB/s/W \\
    softmax GB/s                & 6.1   & 0.51 GB/s/W \\
    layer\_norm GB/s            & 10.9  & 0.76 GB/s/W \\
    gelu light-compute GB/s     & 175.8 & 10.41 GB/s/W \\
    \midrule
    decode tok/s ($B=1\to32$) & 11 / 16 / 32 / 60 / 112 / 194 & fp32 reference; relerr 6e-6 \\
    \bottomrule
  \end{tabular}
\end{adjustbox}
\end{table}

\section{Held-out predictor validation}
\label{sec:predictor-validation}

The energy predictor of the main paper, $E_d(\text{AI})=e_{\text{flop},d}+e_{\text{byte},d}/\text{AI}$, is fitted from the ceiling sweeps alone (ANE weight path $e_\text{flop}=1.16$, $e_\text{byte}=16.5$; GPU $1.66$, $36.3$, all in pJ).
The device-map classes are held out from that fit, so Table~S19 tests the predicted energy magnitude against all of them, grouped by regime.
For single-dispatch workloads the prediction lies below measurement, and near the roof it is within \qtyrange{7}{28}{\percent} (median \qty{14}{\percent} across both devices).
Away from the roof it under-predicts the dispatch-bound rows by large factors, because their package energy is dominated by the fixed per-call cost the roofline omits.
The fused 16-deep conv stack lies outside this domain because it amortizes one dispatch across sixteen layers, and it runs below the single-dispatch floor (\qty{0.44}{} measured against \qty{1.18}{\pico\joule\per\flop} predicted).

\begin{table}[htbp]
  \centering
  \small
  \caption{Held-out predictor validation across the device-map classes (M5 Pro).
  Predicted energy per FLOP $E_d(\text{AI})=e_{\text{flop},d}+e_{\text{byte},d}/\text{AI}$ from the ceiling-fitted constants alone (ANE weight path $e_\text{flop}=1.16$, $e_\text{byte}=16.5$; GPU $1.66$, $36.3$), against the measured value $1000/(\text{perf/W})$ from the main paper's device map (decode from the main paper's decode table).
  The predictor is a one-sided lower bound: tight near the roof, loose in the dispatch-bound regimes where energy is dominated by the fixed per-call cost the roofline omits.
The GPU constant (\qty{36.3}{\pico\joule\per\byte}) is its streaming-roof byte; the GPU has no separate weight path, and the device-pair crossover stays negative, so the ANE remains the lower-energy device, even under the GPU's own cheaper weight-stream byte (about \qty{24}{\pico\joule\per\byte}).
Constants are fitted on the M5 Pro.}
  \label{tab:predictor}
  \footnotesize
  \setlength{\tabcolsep}{4pt}
  \begin{adjustbox}{max width=\linewidth}
\begin{tabular}{llrcc}
    \toprule
    & & & ANE pJ/FLOP & GPU pJ/FLOP \\
    Workload & regime & AI & pred / meas (err) & pred / meas (err) \\
    \midrule
    \multicolumn{5}{l}{\emph{near the roof (bound tight)}}\\
    GEMM compute $K=4096$      & compute-bound  & 227.6 & $1.23$ / $1.33$ ($-7\%$)  & $1.82$ / $2.00$ ($-9\%$) \\
    Attention long-seq $S=512$ & near-roof      & 52.2  & $1.48$ / $1.78$ ($-17\%$) & $2.36$ / $3.28$ ($-28\%$) \\
    Decode loop $B=1$          & weight-roof    & 1.0   & $17.7$ / $22.9$ ($-23\%$) & $38.0$ / $42.5$ ($-11\%$) \\
    \midrule
    \multicolumn{5}{l}{\emph{away from the roof (one-sided bound, loose)}}\\
    GEMM bandwidth $K=1024$    & bandwidth      & 102.4 & $1.32$ / $3.65$ ($-64\%$) & $2.01$ / $6.25$ ($-68\%$) \\
    conv single ($16\to64$)    & dispatch-floor & 224.8 & $1.23$ / $40.8$ ($-97\%$) & $1.82$ / $67.8$ ($-97\%$) \\
    GEMM floor $M=64$          & dispatch-floor & 42.7  & $1.55$ / $63.6$ ($-98\%$) & $2.51$ / $128.7$ ($-98\%$) \\
    \bottomrule
  \end{tabular}
\end{adjustbox}
\end{table}

Both loose regimes are the ones the corrections already identify.

\section{fp16 numerical accuracy per class}
\label{sec:accuracy}

\Cref{tab:accuracy} reports the relative $L_2$ error of the fp16 device path against an fp32/fp64 \texttt{numpy} reference for each device-map class.
Accuracy is set by the fp16 arithmetic and the reference rather than by the host, and the per-class error matches across the M5 Pro, M1 Max, M2 Pro, and M4 to within last-digit measurement noise.
The one large-error class is the compute-bound square GEMM ($K\geq4096$), where fp16 accumulation over a long reduction reaches \num{2.7e-2}.
Every other class stays at \num{1e-3} to \num{1e-2}, the same order as the GPU's fp16.

\begin{table}[htbp]
  \centering
  \footnotesize
  \caption{fp16 relative $L_2$ error per device-map class (M5 Pro; host-invariant to measurement noise).
GPU values are MLX fp16 against the same reference.}
  \label{tab:accuracy}
  \begin{adjustbox}{max width=\linewidth}
\begin{tabular}{lll}
    \toprule
    Class (workload) & ANE relerr & GPU relerr \\
    \midrule
    GEMM floor (M64$\cdot$K256$\cdot$N256)        & 1.3e-3 & 3.6e-4 \\
    GEMM bandwidth (M128$\cdot$K1024$\cdot$N1024) & 5.5e-3 & 3.6e-4 \\
    GEMM compute (M256$\cdot$K4096$\cdot$N4096)   & 2.7e-2 & 3.6e-4 \\
    conv single ($16\to64$)                     & 5.1e-4 & 3.6e-4 \\
    conv ResNet-like ($64\to256$, 16-deep)       & 3.3e-3 & 1.4e-3 \\
    attention ViT block ($S=197$)               & 1.6e-3 & 1.2e-3 \\
    attention long-seq ($S=512$)                & 2.4e-3 & 1.3e-3 \\
    layer\_norm (197,768)                         & 6.7e-4 & 6.1e-4 \\
    rms\_norm (197,768)                           & 5.7e-4 & 5.4e-4 \\
    group\_norm (1,256,$64^2$)                    & 7.4e-4 & 6.0e-4 \\
    5-pt stencil ($256^2$, 32 steps)              & 1.4e-2 & 6.6e-4 \\
    \bottomrule
  \end{tabular}
\end{adjustbox}
\end{table}

\begin{table}[htbp]
  \centering
  \footnotesize
  \caption{The two latency statistics per device-map class (M5 Pro).
\emph{min} is the minimum-latency probe reported in the main paper's device map.
\emph{sust.} is the sustained per-iteration time $\Delta t / n$ measured inside the power window and used for every perf/watt and mJ/inference figure.
The ratio compares the two.
It exceeds $1.25$ only at the two smallest ANE shapes and at the DFT, and stays at $1.01$ to $1.02$ on the three real models.
Substituting the mean package power for the median across these points changes the reported energies by a median of \qty{0.5}{\percent} (range \qtyrange{-5.9}{12.8}{\percent}, under \qty{5}{\percent} at \qty{76}{\percent} of points) and reverses no device comparison.}
  \label{tab:latency-statistics}
  \begin{adjustbox}{max width=\linewidth}
\begin{tabular}{lrrrrrr}
    \toprule
    & \multicolumn{3}{c}{ANE} & \multicolumn{3}{c}{GPU} \\
    \cmidrule(lr){2-4}\cmidrule(lr){5-7}
    Class (workload) & min (ms) & sust.\ (ms) & ratio & min (ms) & sust.\ (ms) & ratio \\
    \midrule
    GEMM floor (M64$\cdot$K256$\cdot$N256) & 0.092 & 0.182 & 1.99 & 0.124 & 0.165 & 1.33 \\
    GEMM bandwidth (M128$\cdot$K1024$\cdot$N1024) & 0.086 & 0.106 & 1.24 & 0.187 & 0.179 & 0.95 \\
    GEMM compute (M256$\cdot$K4096$\cdot$N4096) & 1.293 & 1.314 & 1.02 & 0.461 & 0.521 & 1.13 \\
    conv single $3{\times}3$ & 0.103 & 0.129 & 1.25 & 0.173 & 0.168 & 0.97 \\
    conv ResNet-like $3{\times}3{\times}16$ & 0.757 & 0.795 & 1.05 & 3.159 & 3.215 & 1.02 \\
    attention ViT ($S{=}197$) & 0.230 & 0.236 & 1.03 & 0.269 & 0.333 & 1.24 \\
    attention long-seq ($S{=}512$) & 0.817 & 0.833 & 1.02 & 0.390 & 0.455 & 1.17 \\
    layer\_norm (197,768) & 0.195 & 0.210 & 1.08 & 0.176 & 0.181 & 1.03 \\
    rms\_norm (197,768) & 0.373 & 0.391 & 1.05 & 0.151 & 0.175 & 1.15 \\
    group\_norm (1,256,$64^2$) & 0.350 & 0.366 & 1.05 & 0.196 & 0.251 & 1.28 \\
    DFT-as-matmul ($N{=}1024$) & 0.176 & 0.422 & 2.39 & 0.185 & 0.170 & 0.92 \\
    5-pt stencil ($256^2$, 32 steps) & 0.128 & 0.134 & 1.05 & 1.256 & 1.327 & 1.06 \\
    Jacobi ($n{=}512$, 25 iters) & 0.170 & 0.173 & 1.02 & 0.285 & 0.331 & 1.16 \\
    ResNet-18 forward & 0.330 & 0.337 & 1.02 & -- & -- & -- \\
    MiniLM encoder & 0.525 & 0.536 & 1.02 & -- & -- & -- \\
    ViT-B/16 forward & 18.336 & 18.450 & 1.01 & -- & -- & -- \\
    \bottomrule
  \end{tabular}
\end{adjustbox}
\end{table}

\begin{table}[htbp]
  \centering
  \footnotesize
  \caption{The dispatch ratio $\rho = $ sustained per-iteration time over minimum-latency probe, ANE path, across three generations.
The M5 Pro column is measured directly from the retained iteration counts (\Cref{tab:latency-statistics}).
The M4 and M2 Pro records retain the probe throughput, the active power and the derived perf/watt, but not the sustained time, so $\rho$ is recovered from the identity $\rho = \text{GFLOP/s} / (\text{GFLOP/s/W} \times \text{W})$, which is exact up to the stored precision.
Re-rounding the M5 Pro record to that precision and applying the identity reproduces the measured $\rho$ to a median of \qty{0.04}{\percent} and a worst case of \qty{0.3}{\percent} over 20 operating points, so the recovered columns are limited by the stored precision and not by the method.
The three normalization classes are omitted because their perf/watt is reported in items/s/W, which does not close the identity.}
  \label{tab:dispatch-ratio}
  \begin{adjustbox}{max width=\linewidth}
\begin{tabular}{lrrr}
    \toprule
    & \multicolumn{1}{c}{measured} & \multicolumn{2}{c}{recovered} \\
    \cmidrule(lr){2-2}\cmidrule(lr){3-4}
    Class (workload) & M5 Pro & M4 & M2 Pro \\
    \midrule
    GEMM floor (M64$\cdot$K256$\cdot$N256) & 1.99 & 3.01 & 2.46 \\
    GEMM bandwidth (M128$\cdot$K1024$\cdot$N1024) & 1.24 & 0.88 & 1.23 \\
    GEMM compute (M256$\cdot$K4096$\cdot$N4096) & 1.02 & 1.01 & 0.98 \\
    conv single $3{\times}3$ & 1.25 & 1.01 & 1.35 \\
    conv ResNet-like $3{\times}3{\times}16$ & 1.05 & 1.03 & 1.00 \\
    attention ViT ($S{=}197$) & 1.03 & 0.94 & 1.02 \\
    attention long-seq ($S{=}512$) & 1.02 & 1.00 & 0.98 \\
    DFT-as-matmul ($N{=}1024$) & 2.39 & 0.90 & 0.66 \\
    5-pt stencil ($256^2$, 32 steps) & 1.05 & 1.17 & 1.11 \\
    Jacobi ($n{=}512$, 25 iters) & 1.02 & 1.06 & 1.04 \\
    \bottomrule
  \end{tabular}
\end{adjustbox}
\end{table}

The dispatch floor at the smallest GEMM reproduces on all three generations.
No other class does.
The DFT-as-matmul is strongly dispatch-bound on the M5 Pro alone, and ratios below unity, where the closed loop outruns the single-call probe, appear on both older hosts.
The dispatch term of the main paper's cost model is therefore fitted on the M5 Pro only.

\begin{table}[htbp]
  \centering
  \footnotesize
  \caption{Decode weight encodings at $B=1$ on the configuration of the main paper's decode table (M5 Pro).
A requested encoding is not always the applied one, so the weight blob is read from the compiled artifact rather than inferred from the compile flag, and the emitted operator is read from the generated MIL.
\texttt{int4} at the default reconstruction-error budget of $0.05$ is rejected for every weight of this model and falls back silently to per-channel int8, which the \texttt{int4}$^\dagger$ row records.
Latencies are means over five interleaved rounds of 30 repetitions, round-robin across encodings.
Across-round standard deviations are \qtyrange{0.03}{0.3}{\percent} of the mean.}
  \label{tab:decode-quant}
  \begin{adjustbox}{max width=\linewidth}
\begin{tabular}{lrlrrrr}
    \toprule
    Encoding & weight blob & emitted operator & lat (ms) & tok/s & GB/s & Gelem/s \\
    \midrule
    fp16 & \qty{1149.1}{\mega\byte} & (dense \texttt{const}) & 8.598 & 116.3 & 140.1 & 70.1 \\
    int8 & \qty{575.0}{\mega\byte} & \texttt{affine\_dequantize} & 8.100 & 123.5 & 74.4 & 74.4 \\
    int4$^\dagger$ & \qty{575.0}{\mega\byte} & \texttt{affine\_dequantize} & 8.114 & 123.2 & 74.3 & 74.2 \\
    int4 & \qty{287.3}{\mega\byte} & \texttt{lut\_to\_dense} & 4.923 & 203.1 & 61.2 & 122.4 \\
    blockwise & \qty{610.5}{\mega\byte} & \texttt{blockwise\_shift\_scale} & 8.892 & 112.5 & 72.0 & 67.7 \\
    \bottomrule
  \end{tabular}
\end{adjustbox}

  \vspace{4pt}
  {\footnotesize $^\dagger$ requested int4 at the default budget; applied encoding is int8.}
\end{table}

Engagement is a binary switch on the reconstruction-error budget, not a continuum.
At the default $0.05$ every weight of this model is rejected and the encoding falls back to int8.
From $0.15$ upward the encoder emits the same palette at every budget we tested: sweeping $0.15$, $0.25$, $0.40$ and $0.60$ gives one weight blob (\qty{301283360}{\byte}) and one relative $L_2$ (\num{0.5766}) at all four points, with latency varying by \qty{0.3}{\percent}.
That sweep therefore fixes where int4 engages, and it does not vary the palette's accuracy, so it cannot show whether speed tracks accuracy.

The relative $L_2$ is measured against an fp64 reference on randomly initialized weights, which are unfavorable for 4-bit palettization because a Gaussian gives the palette no cluster structure to exploit.
The evidence that the LUT path streams every weight is \Cref{tab:imagenet} rather than this error figure.
Requested at the default budget on ResNet-18 the encoding falls back and reproduces the int8 row exactly (\qty{69.65}{\percent} top-1 and \qty{2.33}{\percent} disagreement in both rows).
Engaged, it is a distinct and working classifier at \qty{67.21}{\percent} top-1 over all \num{50000} validation images, which a path that skipped weights could not reach.

\begin{table}[htbp]
  \centering
  \footnotesize
  \caption{Decode against KV-cache length at $B=1$ (M5 Pro), the axis the main paper's decode table holds fixed at 256.
In this sweep the GPU cache is resident while the ANE cache is re-fed as a graph input on every step, so the last column is not a like-for-like device comparison.
It is retained for comparison with \Cref{tab:kv-residency}.}
  \label{tab:decode-context}
  \begin{adjustbox}{max width=\linewidth}
\begin{tabular}{rrrrrrrr}
    \toprule
    & \multicolumn{3}{c}{ANE} & \multicolumn{2}{c}{GPU} & \multicolumn{2}{c}{ratio} \\
    \cmidrule(lr){2-4}\cmidrule(lr){5-6}\cmidrule(lr){7-8}
    $s_{kv}$ & lat (ms) & tok/s & rail (mW) & lat (ms) & tok/s & GPU/ANE tok/s & ANE/GPU tok/s/W \\
    \midrule
    128  & 8.238  & 121.4 & 1906 & 4.871 & 205.3 & 1.69 & 1.66 \\
    256  & 8.605  & 116.2 & 1876 & 4.940 & 202.4 & 1.74 & 1.69 \\
    512  & 9.482  & 105.5 & 1822 & 5.116 & 195.4 & 1.85 & 1.56 \\
    1024 & 11.411 & 87.6  & 1644 & 5.499 & 181.8 & 2.07 & 1.10 \\
    2048 & 14.885 & 67.2  & 1431 & 6.205 & 161.2 & 2.40 & 0.81 \\
    \bottomrule
  \end{tabular}
\end{adjustbox}
\end{table}

Per-step latency is linear in context on both devices.
The ANE fit is \qty{7.750}{\milli\second} plus \qty{3.496}{\micro\second} per KV position and the GPU fit \qty{4.769}{\milli\second} plus \qty{0.702}{\micro\second}, with maximum residuals of \qty{0.71}{\percent} and \qty{0.24}{\percent}.
The intercepts recover the batch-invariant weight stream fitted from the batch sweep, and the KV term reaches the weight term only at $s_{kv}\approx2200$ on the ANE and $\approx6800$ on the GPU.
Re-feeding costs the ANE a host-side copy that reaches \qty{128}{\mega\byte} per token, which is why the efficiency ratio in the last column falls to $0.81$ at 2048.
Held resident on both devices, the ANE's advantage is \numrange{2.6}{2.0} across the same sweep with no crossover, and its fitted per-position slope drops from \qty{3.49}{} to \qty{1.34}{\micro\second}.
The weight-stream intercept is unaffected by the transport, so the linearity and the intercept recovery stand either way.

\begin{table}[htbp]
  \centering
  \footnotesize
  \caption{Task-level accuracy on the ILSVRC2012 validation set for the two classifiers (M5 Pro).
Every device is given the identical preprocessed tensor, so the only difference between rows is arithmetic and weight encoding, and the ANE loader carries the same torchvision weights as the reference.
\emph{Disagree} is the fraction of images whose predicted label differs from the fp32 reference.
\emph{b} counts images the reference gets right and the device gets wrong, \emph{c} the reverse, and $p$ is McNemar's exact test on those discordant pairs, which tests whether the reduced precision changes decisions in a direction.
The int4$^\dagger$ row is the default accuracy budget, which falls back to int8 for every weight of this model and so reproduces the int8 row exactly.
ViT uses \num{20000} images and not all \num{50000} because its fp32 CPU reference costs \qty{35.7}{\milli\second} per image.}
  \label{tab:imagenet}
  \begin{adjustbox}{max width=\linewidth}
\begin{tabular}{lrrrrrr}
    \toprule
    Device & top-1 & \qty{95}{\percent} CI & top-5 & disagree & $b/c$ & McNemar $p$ \\
    \midrule
    \multicolumn{7}{l}{\emph{ResNet-18, all \num{50000} validation images}} \\
    CPU fp32 (reference) & \qty{69.76}{\percent} & [69.36, 70.16] & \qty{89.08}{\percent} & --- & --- & --- \\
    ANE fp16 & \qty{69.76}{\percent} & [69.36, 70.16] & \qty{89.09}{\percent} & \qty{0.11}{\percent} & 14/14 & 1.00 \\
    GPU MPS fp16 & \qty{69.76}{\percent} & [69.36, 70.16] & \qty{89.10}{\percent} & \qty{0.14}{\percent} & 16/17 & 1.00 \\
    ANE int8 & \qty{69.65}{\percent} & [69.24, 70.05] & \qty{88.97}{\percent} & \qty{2.33}{\percent} & 319/263 & 0.022 \\
    ANE int4$^\dagger$ & \qty{69.65}{\percent} & [69.24, 70.05] & \qty{88.97}{\percent} & \qty{2.33}{\percent} & 319/263 & 0.022 \\
    ANE int4 & \qty{67.21}{\percent} & [66.79, 67.62] & \qty{87.89}{\percent} & \qty{13.89}{\percent} & 2514/1237 & $<10^{-16}$ \\
    \midrule
    \multicolumn{7}{l}{\emph{ViT-B/16, \num{20000} validation images}} \\
    CPU fp32 (reference) & \qty{81.06}{\percent} & [80.51, 81.60] & \qty{95.38}{\percent} & --- & --- & --- \\
    ANE fp16 & \qty{81.00}{\percent} & [80.45, 81.54] & \qty{95.35}{\percent} & \qty{0.40}{\percent} & 33/21 & 0.134 \\
    GPU MPS fp16 & \qty{81.07}{\percent} & [80.52, 81.61] & \qty{95.38}{\percent} & \qty{0.04}{\percent} & 3/5 & 0.727 \\
    \bottomrule
  \end{tabular}
\end{adjustbox}

  \vspace{4pt}
  {\footnotesize $^\dagger$ requested int4 at the default budget; applied encoding is int8.}
\end{table}

Both fp32 references reproduce the published torchvision top-1 (\qty{69.758}{} and \qty{81.072}{\percent}), which validates the preprocessing and label mapping.
McNemar finds no directional change for either fp16 path.
ViT-B/16 has the largest ANE relative $L_2$ of the paper's three real models, \num{2.6e-2} on real images against the \num{7.8e-3} a random-input measurement gives.
Even there the ANE's top-1 is statistically indistinguishable from fp32, and although it changes ten times as many individual labels as the GPU does ($0.40$ against \qty{0.04}{\percent}), those changes are not directional.

\begin{table}[htbp]
  \centering
  \footnotesize
  \caption{Task-level check for the sentence encoder (MiniLM, \num{4977} wikitext-2-raw sentences, M5 Pro).
For this embedding model, task-level agreement is measured by nearest-neighbor retrieval.
Each device embeds the same corpus.
\emph{top-1 NN} is the fraction of sentences whose nearest neighbor under cosine similarity is unchanged from the fp32 reference, \emph{top-10} is the mean overlap of the ten nearest, and the last column is the Spearman correlation of the pairwise similarities themselves.
These values measure agreement with fp32, not accuracy against ground truth, because no labeled sentence-pair set was available offline.}
  \label{tab:retrieval}
  \begin{adjustbox}{max width=\linewidth}
\begin{tabular}{lrrrr}
    \toprule
    Device & rel.\ $L_2$ & top-1 NN & top-10 overlap & similarity Spearman \\
    \midrule
    CPU fp32 (reference) & --- & \qty{100.00}{\percent} & \qty{100.00}{\percent} & $1.0000$ \\
    ANE fp16 & \num{5.7e-3} & \qty{99.18}{\percent} & \qty{99.39}{\percent} & $1.0000$ \\
    GPU MPS fp16 & \num{1.3e-3} & \qty{99.78}{\percent} & \qty{99.85}{\percent} & $1.0000$ \\
    ANE int8 & \num{3.1e-2} & \qty{97.31}{\percent} & \qty{97.84}{\percent} & $0.9994$ \\
    \bottomrule
  \end{tabular}
\end{adjustbox}
\end{table}

The ordering matches the classifiers.
fp16 on either device preserves the retrieved neighbor for more than \qty{99}{\percent} of queries and the similarity structure to four decimals, and per-channel int8 costs about two points of retrieval agreement.
For the two classifiers, where both agreement and accuracy are measurable, they give the same verdict.

\paragraph{Operators excluded from the speed comparison.} Sixteen operators reach the engine through a one-shot subprocess bridge, paying the full process-spawn, graph-author and ANE-load cost on every call, so timing them against MLX would measure the bridge overhead (main paper, Sec.~2).
\begin{sloppypar}
They are \texttt{sort}, \texttt{cross\_product}, \texttt{cross\_cor\-relation}, \texttt{cost\_volume}, \texttt{fps},
\texttt{radius\_search}, \texttt{minmax\_norm}, \texttt{lrn}, \texttt{space\_to\_channel}, \texttt{channel\_to\_space},
\texttt{space\_to\_batch}, \texttt{batch\_to\_space}, \texttt{flatten}, \texttt{input\_view}, \texttt{dynamic\_slice}
and \texttt{scaled\_elementwise}.
\end{sloppypar}
The three bridge operators that carry a persistent worker (\texttt{sdpa}, \texttt{argmax}, \texttt{topk}) are timed on silicon and stay in the comparison.

\paragraph{Scope of the relative-error column.} The per-class relative $L_2$ reported throughout uses seeded Gaussian input.
It does not consistently bound the error on real data or preserve the ranking across models.
Re-measuring the three real models on their own domains, ResNet-18 falls from \num{2.3e-3} to \num{1.6e-3}, MiniLM holds at \num{5.7e-3} against a reported \num{5.5e-3}, and ViT-B/16 rises from \num{7.8e-3} to \num{2.6e-2}.

ViT-B/16 permits a layerwise decomposition of this discrepancy.
We compile encoder prefixes of depth 1 to 12 and compare each against the corresponding fp32 reference.
Error grows by a median factor of $1.07$ to $1.09$ per block for both input types, and through depth 11 the real-over-synthetic ratio stays between $0.85$ and $1.30$.
The entire discrepancy appears at the last block and the classification head.
There the absolute error on real images jumps to $2.24\times$ the synthetic figure while the reference logit vector's RMS falls to $0.67\times$ of it, and their ratio, $3.34$, reproduces the observed $3.33\times$ ratio in relative $L_2$.
Both terms move the same way because a trained classifier on a real image produces a peaked logit vector.
A few large entries carry larger absolute fp16 rounding, while the vector's RMS over all 1000 classes is lower than for the diffuse response to noise.
Relative $L_2$ over a logit vector is therefore least informative for confident predictions, the regime that decides the class.
The encoder shows ordinary accumulation of fp16 rounding rather than a single-operator failure.

\begin{table}[htbp]
  \centering
  \footnotesize
  \caption{KV-cache residency and its cost (M5 Pro, $B=1$, TinyLlama-class decoder of \Cref{tab:decode-context}).
The GPU path holds the cache resident in device arrays across steps.
The ANE path can either re-feed it as a graph input every step or hold it in the engine's own input buffers and re-run without rebinding.
Both produce identical logits (relative $L_2$ exactly zero), so the difference is transport and not arithmetic.
Re-feeding charges the ANE a host-side copy the GPU never pays, and at $s_{kv}=2048$ that copy is \qty{128}{\mega\byte} per token.}
  \label{tab:kv-residency}
  \begin{adjustbox}{max width=\linewidth}
\begin{tabular}{rlrrrrr}
    \toprule
    $s_{kv}$ & arm & lat (ms) & pkg (W) & ANE rail (mW) & CPU rail (mW) & ANE/GPU tok/s/W \\
    \midrule
    256 & ANE re-fed & 8.600 & 3.75 & 1878 & 1730 & 1.79 \\
        & ANE resident & 8.089 & 2.83 & 1950 & 884 & \textbf{2.56} \\
        & GPU & 4.918 & 11.42 & --- & 4692 & --- \\
    \midrule
    1024 & ANE re-fed & 11.384 & 4.96 & 1640 & 3316 & 1.09 \\
         & ANE resident & 9.124 & 2.60 & 1950 & 652 & \textbf{2.63} \\
         & GPU & 5.499 & 11.15 & --- & 4036 & --- \\
    \midrule
    2048 & ANE re-fed & 14.872 & 5.86 & 1428 & 4428 & 0.87 \\
         & ANE resident & 10.492 & 3.68 & 1915 & 1740 & \textbf{1.97} \\
         & GPU & 6.223 & 12.09 & --- & 4563 & --- \\
    \bottomrule
  \end{tabular}
\end{adjustbox}
\end{table}

The rail columns separate the two transports.
Re-fed, the ANE rail decreases with context (\qtyrange{1878}{1428}{\milli\watt}) while package power rises, because the engine waits on the host.
Resident, the rail is flat between \qty{1915}{} and \qty{1950}{\milli\watt} and package power is lower.
The efficiency ratio turns on this choice alone, from an apparent GPU advantage at long context to a steady ANE advantage of about $2\times$.

\begin{table}[htbp]
  \centering
  \footnotesize
  \caption{The DRAM rail, measured on the IOReport \texttt{Energy Model} DRAM channel across four independent runs of the estimator cross-validation (M5 Pro; idle-subtracted active power).
Package here is the ANE${+}$GPU${+}$CPU total that \texttt{powermetrics} reports as Combined Power, which excludes DRAM, so the last column is the fraction the main metric omits.}
  \label{tab:dram}
  \begin{adjustbox}{max width=\linewidth}
\begin{tabular}{llll}
    \toprule
    Operating point & DRAM W (4 runs) & Package W (4 runs) & DRAM \% of package \\
    \midrule
    idle & 0.038 / 0.076 / 0.771 / 0.076 & 0.34 / 0.80 / 5.22 / 0.98 & 7.7--14.8 \\
    ANE conv stack & 0.238 / 0.360 / 0.307 / 0.256 & 5.39 / 6.77 / 7.03 / 6.33 & 4.1--5.3 \\
    ANE attention & 0.811 / 1.127 / 0.868 / 0.790 & 8.16 / 12.27 / 8.69 / 8.45 & 9.2--10.0 \\
    ANE weight stream & 0.934 / 1.061 / 1.024 / 0.893 & 8.34 / 9.33 / 9.63 / 8.48 & 10.5--11.4 \\
    GPU GEMM (saturating) & 6.719 / 7.112 / 6.490 / 6.646 & 53.18 / 54.99 / 51.82 / 52.87 & 12.5--12.9 \\
    CPU GEMM & 0.406 / 0.657 / 1.170 / 0.393 & 16.68 / 18.29 / 19.26 / 16.46 & 2.4--6.1 \\
    \bottomrule
  \end{tabular}
\end{adjustbox}
\end{table}

\paragraph{The latency distribution.} The probe used throughout the paper reports the minimum wall time over 30 repetitions after warmup; here we also give the median and the mean.
We re-ran the probe over the device-map workloads with the full distribution retained, using the same workload builders so that shapes, dtypes and seeds are identical by construction.
Across 46 device-and-workload probes the median exceeds the minimum by a median factor of $1.086$ (mean $1.117$, p90 $1.178$), and \qty{85}{\percent} of probes lie within \qty{25}{\percent} of their minimum.
Every probe whose median exceeds its minimum by more than \qty{30}{\percent} is a GPU or CPU path.
\texttt{group\_norm} on the GPU is the extreme at $2.51$ (per-repetition CV \qty{41}{\percent}), followed by GPU attention at $1.58$ and GPU convolution at $1.57$.
The ANE paths sit between $1.02$ and $1.09$ on every class.
Minimum latency benefits the more variable GPU and CPU paths more strongly, so using medians would increase the reported ANE/GPU ratios.

\section{Formulation fairness of the GPU baselines}
\label{sec:formulation}

Kernel fusion and workload formulation are separate confounds.
The fused-GPU appendix (\Cref{sec:fused-gpu}) tests the first and finds that MLX kernel-launch overhead does not explain the device map's large gaps.
This appendix audits formulation and remeasures the stencil and the DFT.

\paragraph{Formulation-exposed classes.} A class is formulation-exposed when the GPU's idiomatic expression discards a better algorithm for the same problem.
The GEMM, convolution, attention and norm classes are not.
Matrix multiply, \texttt{conv2d} and elementwise reductions are the standard formulations tested here, and the fused-GPU appendix additionally checked convolution, attention and two norms under \texttt{mx.compile}.
The Jacobi solve is a dense $\mathbf{R}\mathbf{x}$ recurrence, which is likewise standard.
The stencil and the DFT both have known lower-work formulations.
Two classes are exposed.
The 5-point stencil is expressed as a dense $3\times3$ convolution whose kernel carries four structural zeros, so the GPU reads nine neighbors to use five.
The DFT is expressed as two dense $N\times N$ matrix--vector products, where a fast transform is $O(N\log N)$.

\paragraph{The formulation-bound stencil.} Rewriting the GPU stencil as direct shifted-slice arithmetic on a zero-padded field, $u_{n+1} = u_n + \Delta t\,(u_{i-1,j} + u_{i+1,j} + u_{i,j-1} + u_{i,j+1} - 4u_{n})$, and compiling it with \texttt{mx.compile}, gives \qty{0.382}{\milli\second} against the dense convolution's \qty{1.256}{\milli\second} and \qty{11.6}{\watt} against \qty{33.4}{\watt}.
The rewrite is $3.3\times$ faster and draws $2.9\times$ less power, and it is also slightly more accurate (relative $L_2$ \num{5.8e-4} against \num{6.6e-4}, both against the same fp64 reference).
The optimized baseline preserves the direction but reduces the perf/watt ratio by $9.3\times$, from $43.8\times$ (95\% CI $[35.6, 52.0]$, $N{=}5$) to $4.71\times$ ($[3.94, 5.49]$), and the latency advantage from $9.2\times$ to $2.8\times$.
The main paper's device map reports the optimized baseline.

Both devices are close to their dispatch floors at this size.
A linear fit to the GPU's step-count sweep gives \qty{5.87}{\micro\second} per step plus \qty{0.184}{\milli\second} fixed, so \qty{49}{\percent} of the GPU time at 32 steps is fixed cost, and \texttt{aneforge}'s own cost model flags the ANE program as dispatch-floor-bound.
The comparison is asymmetric in the GPU's favor: it runs hand-written code while the ANE runs the framework's natural expression.
Convolution is the ANE's native primitive, it has no 5-tap operator, and a shift formulation would place the work on the slow activation path, so we take the comparison to be between each device's best available expression.

\paragraph{The DFT's unresolved verdict.} Replacing the dense matrix--vector products with \texttt{mx.fft} is only $1.17\times$ faster (\qty{0.172}{\milli\second} against \qty{0.201}{\milli\second}) despite an $82\times$ smaller operation count, because at $N=1024$ every variant is dispatch-bound, not compute-bound.
The FFT is, however, \num{1.2e3} times more accurate (relative $L_2$ \num{2.9e-7} against \num{3.5e-4}), which is worth stating because this class was chosen as an fp16 cancellation stress and much of that stress is an artifact of the formulation.

Over ten independent cold starts the ANE-over-GPU perf/watt ratio is $1.35$ with a 95\% CI of $[0.96, 1.73]$, a paired-$t$ $p$ of $0.073$, and a TOST equivalence $p$ of $0.93$, so neither a difference nor an equivalence is established.
The minimum detectable effect at this $N$ is \qty{35}{\percent} of the pooled mean, and the device map's single-session ratio ($0.82\times$, the GPU ahead) lies outside the interval.
The cause is the power scale.
This class draws \qtyrange{3}{9}{\watt} on the ANE while the idle baseline being subtracted varied over \qtyrange{156}{632}{\milli\watt} across cold starts, so the subtraction is a large and drifting fraction of the signal.
The DFT result is therefore indeterminate.

\paragraph{The remaining GPU-favorable class.} Applying the same protocol to \texttt{group\_norm}, the only remaining class where the GPU leads, gives a ratio of $0.949$ with a 95\% CI of $[0.900, 0.997]$ over ten cold starts (paired-$t$ $p=0.035$, minimum detectable effect \qty{7}{\percent} of pooled), reproducing the single-session $0.943$.
That class draws about \qty{10}{\watt}, where the idle drift is a small fraction of the signal.
The contrast between the two rows locates the instrument's working range, not a property of either workload.

\paragraph{Real models.} The real-model rows were first measured on torch-MPS, the one framework that differs from the rest of the paper.
Re-measuring ResNet-18 against a graph-compiled MLX baseline gives \qty{1.645}{\milli\second} and \qty{32.6}{\milli\joule} per inference, against torch-MPS at \qty{2.351}{\milli\second} and \qty{21.1}{\milli\joule}, so compiled MLX is \qty{30}{\percent} faster while drawing $2.2\times$ the power and using more energy per inference.
The ANE's energy advantage on this model is $14.6\times$ (95\% CI $[12.1, 17.2]$, $N{=}5$) against compiled MLX and $8.9\times$ ($[7.5, 10.7]$, $N{=}16$) against torch-MPS.
The torch-MPS ResNet-18 energy spans \qtyrange{20}{49}{\milli\joule} over independent cold starts, a coefficient of variation of \qty{30}{\percent} against \qty{11}{\percent} for the ANE, so most of the uncertainty in the ratio comes from the baseline and not the device under test.

\paragraph{MiniLM and ViT-B/16.} The same check on the two transformers needs an MLX implementation of each, and the MLX image library used for ResNet-18 has neither.
We ported both to plain MLX, loading the identical sentence-transformers and torchvision weights and using MLX's fused attention and layer-norm kernels (\texttt{mx.fast.\allowbreak scaled\_dot\_product\_\allowbreak attention} and \texttt{mx.fast.layer\_norm}).
Run in fp32 on the MLX CPU stream, the ports agree with the torch fp32 references to a relative error of \num{6.5e-7} (MiniLM) and \num{4.6e-7} (ViT-B/16).
In fp16 on the GPU their errors are \num{1.4e-3} and \num{1.3e-3}, and compilation leaves the output bit-identical.
The four arms (ANE, torch-MPS fp16, eager MLX fp16, compiled MLX fp16) were interleaved within one session for ten repetitions (MLX~v0.32.2).

On MiniLM, compiled MLX takes \qty{0.650}{\milli\second} and \qty{11.5}{\milli\joule} per inference against the ANE's \qty{0.529}{\milli\second} and \qty{3.49}{\milli\joule}, an ANE energy advantage of $3.42\times$ (95\% CI $[2.99, 3.86]$).
On ViT-B/16, compiled MLX takes \qty{2.96}{\milli\second} and \qty{100.4}{\milli\joule}, against torch-MPS fp16 at \qty{15.8}{\milli\second} and \qty{598}{\milli\joule}.
The ANE took \qty{18.8}{\milli\second} in this session, $6.4\times$ the compiled-MLX time ($6.2\times$ against the probe latency of the main paper's Table~6), and its energy advantage falls to $1.53\times$ ($[1.26, 1.81]$).
One ANE repetition read \qty{199}{\milli\joule} at \qty{10.4}{\watt}, against \qtyrange{56}{77}{\milli\joule} for the other nine.
We kept it in; dropping the three ANE windows the harness flags as low-confidence gives $1.71\times$ ($[1.66, 1.76]$).
The torch-MPS ratios from the same session, $6.58\times$ on MiniLM and $9.12\times$ on ViT-B/16, match the cold-start values of \Cref{tab:real-fp16}.
Compiled MLX is the faster GPU baseline on all three models, but it uses less energy than torch-MPS only on the two transformers.
Result files record the framework versions alongside each measurement.

\paragraph{Per-operator energy per FLOP.} The predictor uses one compute-energy constant per device, fit from the saturating GEMM and convolution sweeps.
That quantity is operator-dependent, so we report the spread here rather than leave the constant unqualified.
At the device-map operating points the ANE spends \qty{1.33}{\pico\joule\per\flop} on saturating GEMM, \qty{1.78}{} and \qty{1.91}{} on the two attention shapes, and \qty{0.44}{} on the fused 16-deep convolution stack.
The GPU spends \qty{2.00}{}, \qty{3.28}{}, \qty{5.08}{} and \qty{5.71}{} on the same four.
The roughly fourfold spread on each device does not call for per-class constants, because the model is not a per-class estimator.
The fitted constant characterizes single-dispatch work.
Deep fusion can fall below it by amortizing the per-call cost across layers, as the 16-deep stack does at \qty{0.44}{} against the fitted \qty{1.16}{\pico\joule\per\flop}, for the same reason that puts those points above the compute envelope.

\paragraph{The int8 decode agreement protocol.} The greedy-token agreement quoted for int8 decoding is computed on a synthetic 8-layer decoder ($d_{\text{model}}=2048$, 16 heads, $d_{\text{ff}}=8192$, vocabulary \num{32000}, KV length 256) with randomly initialized weights and a random batch of 64 decode positions.
For each position we take the argmax over the \num{32000} logits and compare against an fp64 reference of the same decode, so \qty{92}{\percent} is 59 of 64 positions, alongside a top-5 overlap of \qty{93}{\percent} and a softmax KL of \num{1.3e-3}.
The measurement is one of numerical fidelity on random weights, not of task quality on a trained model.
The fp16 baseline agrees with fp64 on \qty{95}{\percent} of the same positions, so most of the \qty{8}{\percent} int8 disagreement is already present in fp16.

\section{The measurement instrument}
\label{sec:instrument}

Every measurement in the main paper is taken through \texttt{aneforge}, the CoreML-free fused-graph dispatch path documented in the ANEForge software paper~\cite{aneforge_paper}, over the runtime stack described in the companion guide~\cite{ane_guide}.
The software paper establishes how the path reaches the engine.
This section states the guarantees and limitations of that path as a measurement instrument.

\texttt{aneforge} requests direct ANE execution without CoreML placement.
Rail power, compiler and runtime structure, \texttt{MLComputePlan}, and correctness tests support that placement, but the platform exposes no per-operator rail-exclusive trace, so these are necessary-condition checks and not a direct placement readout.

\texttt{aneforge} accepts a user operator graph and lowers it to a single fused program in Apple's Model Intermediate Language (MIL), with one MIL operation emitted per graph node and all weights packed into one constant blob.
It compiles that program with Apple's ANE compiler (\texttt{ANECCompile}) and submits it through Apple's \texttt{e5rt} execution runtime directly to the engine.
There is no CoreML model and therefore no CoreML placement heuristic deciding which operations run where.
Measurements are taken under a chosen batch size, fp16 weights and activations throughout, and warm program and cache state.

The first check is the per-rail power attribution of the main paper's methodology.
\texttt{powermetrics} reports the ANE rail separately, and an ANE workload draws measurable power on that rail while it runs, which a CPU or GPU fallback would not produce.

The second check is structural and ships with the artifact.
The route and capability gate compiles both lowerings of every selectable operator on the device and asserts they agree within fp16 operation noise, which verifies the set of operators the instrument claims to reach.
The correctness corpus runs each shipped operator against a reference, so a node that failed to lower or executed incorrectly would surface as a test failure rather than as a fast but wrong measurement.

The third is a first-party triangulation.
Querying Apple's own planner through \texttt{MLComputePlan} for single-operator fp16 programs, it prefers the Neural Engine for matmul and convolution and the CPU for standalone elementwise operators, the same compute-bound-favors-ANE split the device map reports.
It confirms placement, not the per-engine energy CoreML does not expose.

\clearpage

\end{document}